\documentclass[traditabstract]{aa}

\usepackage[colorlinks=true, linkcolor=blue, citecolor=blue, urlcolor=blue]{hyperref}
\usepackage{amsmath}
\usepackage{graphicx}
\usepackage{txfonts}
\usepackage{amssymb}
\usepackage[]{natbib}
\usepackage{mathrsfs}
\usepackage{caption}
\usepackage{float}
\usepackage{afterpage}
\usepackage{amstext}
\begin{document}
\title{HST/STIS transmission spectrum of the ultra-hot Jupiter WASP-76~b confirms the presence of sodium in its atmosphere}
   \author{C. von Essen$^{1,2}$, M. Mallonn$^3$, S. Hermansen$^{1}$, M. C. Nixon$^4$, N. Madhusudhan$^4$, H. Kjeldsen$^{1,2}$, G. Tautvai{\v s}ien{\. e}$^2$}
   \authorrunning{C. von Essen (2019)}
   \titlerunning{Transmission spectrum of WASP-76~b}
   \institute{$^1$Stellar Astrophysics Centre, Department of Physics and Astronomy, Aarhus University, Ny Munkegade 120, DK-8000 Aarhus C, Denmark\\ 
              $^2$Astronomical Observatory, Institute of Theoretical Physics and Astronomy, Vilnius University, Sauletekio av. 3, 10257, Vilnius, Lithuania\\
              $^3$Leibniz-Institut f\"{u}r Astrophysik Potsdam (AIP), An der Sternwarte 16, D-14482 Potsdam, Germany\\
              $^4$Institute of Astronomy, University of Cambridge, Madingley Road, Cambridge, CB3 0HA, UK\\
     \email{cessen@phys.au.dk}
   }

   \date{Received: 22.11.2019; accepted: 13.03.2020}

\abstract{We present an atmospheric transmission spectrum of the
  ultra-hot Jupiter WASP-76~b by analyzing archival data obtained with
  the Space Telescope Imaging Spectrograph (STIS) on board the Hubble
  Space Telescope (HST). The dataset spans three transits, two with a
  wavelength coverage between 2900 and 5700~\AA, and the third one
  between 5250 and 10300~\AA. From the one-dimensional, time dependent
  spectra we constructed white and chromatic light curves, the latter
  with typical integration band widths of $\sim$200~\AA. We computed
  the wavelength dependent planet-to-star radii ratios taking into
  consideration WASP-76's companion. The resulting transmission
  spectrum of WASP-76~b is dominated by a spectral slope of increasing
  opacity towards shorter wavelengths of amplitude of about three
  scale heights under the assumption of planetary equilibrium
  temperature. If the slope is caused by Rayleigh scattering, we
  derive a lower limit to the temperature of \mbox{$\sim$870
    K}. Following-up on previous detection of atomic sodium derived
  from high resolution spectra, we re-analyzed HST data using narrower
  bands centered around sodium. From an atmospheric retrieval of this
  transmission spectrum, we report evidence of sodium at 2.9$\sigma$
  significance. In this case, the retrieved temperature at the top of
  the atmosphere ($10^{-5}$ bar) is $2300^{+412}_{-392}$ K. We also
  find marginal evidence for titanium hydride. However, additional
  high resolution ground-based data are required to confirm this
  discovery.}

          
   \maketitle

\section{Introduction}

Transiting exoplanets allow us to study their atmospheres and determine their physical and chemical properties through transmission spectroscopy, and to do so we analyze the light of their host stars filtered through their atmospheres. This technique has been used to detect several atomic and molecular species in exoplanetary atmospheres, as for example sodium on \mbox{HAT-P-1~b} \citep{nikolov14}, sodium and potassium on WASP-103~b \citep{Lendl2017}, WASP-39~b \citep{fischer} and WASP-127~b \citep{Chen2018}, H$_2$O and CO on \mbox{HD 209458~b} \citep{deming}, H$_2$O on WASP-19~b \citep{huitson} and HAT-P-32~b \citep{damiano}, VO on WASP-121~b \citep{evans2018}, and clouds and hazes on \mbox{HD 189733~b} \citep{sing11}, HAT-P-18~b \citep{Kirk2017}, WASP-49~b \citep{Lendl2016}, and WASP-6~b \citep{nikolov15}. It was \cite{charbonneau} who first observed an exoplanetary atmosphere on \mbox{HD 209458~b} via the detection of atmospheric neutral sodium at 5980~\AA\ with the Space Telescope Imaging Spectrograph (STIS) mounted on the Hubble Space Telescope (HST). Since then, transmission spectroscopy, space-based and ground-based, has been widely used to characterize exo-atmospheres \citep[see e.g.,][for comparative studies of large samples of exoplanets]{sing16,tsiaras}. 

Ultra-hot Jupiters are a population of planets that have day-side temperatures higher than \mbox{$\sim$2200 K} and a large thermal difference between their hemispheres, thereby offering excellent conditions for detailed studies of the physics and chemistry of their atmospheres \citep{parmentier2018}. Due to the extreme atmospheric temperatures these exoplanets are expected to be more diverse when compared to hot Jupiters. So far, ultra-hot Jupiters that were discovered by transit surveys presenting inflated radii are more likely to be found orbiting hot stars \citep[see e.g.,][]{Hartman2016}. 
It was with the interest in knowing more about these worlds that \cite{west2016} first announced and further examined \mbox{WASP-76~b}, an inflated ultra-hot Jupiter orbiting each $\sim$1.8 days a main sequence star of \mbox{$\sim$6200 K}. The optical transit depth of WASP-76~b is \mbox{$\sim$1.2\%}, but if the planet was not inflated it would merely be $\sim$0.36\%. This translates into a large scale height, of approximately 1250 km above the nominal planet radius. Comparing WASP-76~b to other exoplanets in their mass versus incident flux, it is clear that WASP-76~b has a higher incident flux than other exoplanets of the same mass, which could in principle explain its inflated radius.

The ultra-hot Jupiters that were studied to a good level of detail so far seem to show a large spread of features. For example, WASP-12~b shows a Rayleigh slope with an amplitude of $\sim$2 scale heights \citep{Stevenson2014}. \cite{Hoeijmakers2019} found heavy metals in one of the hottest exoplanets known to date, KELT-9~b. The controversial WASP-19~b shows either titanium oxide \citep{Sedaghati2017} or a flat spectrum \citep{Espinoza2019}. Aluminium oxide was discovered in the atmosphere of WASP-33~b \citep{wasp33b}, and WASP-121~b revealed magnesium and iron absorption at UV wavelengths, absorption of a currently unknown source at short optical wavelengths and indications for VO \citep{evans2018}. \cite{Lendl2017} detected sodium and potassium on WASP-103~b, and interpreted its atmosphere as potentially cloud-free. A recent study carried out by \cite{seidel2019,Zak2019} report on the detection of neutral sodium in the atmosphere of WASP-76~b, obtained analyzing high resolution spectra. In particular, \cite{seidel2019} found the sodium lines to be significantly broadened, and they speculate this broadening to be an indicator of super-rotation in the upper atmosphere of WASP-76~b. 

In this work, we present the characterization of the atmosphere of WASP-76~b from the near UV to the near IR, obtained analyzing three primary transits using archival data of HST/STIS. This article is structured as follows. In Section~\ref{sec_Data_Reduction} we present the observations and the data reduction, and then give a detailed description of the model parameters and fitting process in Section~\ref{sec_dat_mod_fit_param}. In Section~\ref{sec_results} we show our results on the transmission spectrum of WASP-76~b, we determine the impact of third light contamination into our transmission spectrum in Section~\ref{sec:TLC}, we investigate some mechanisms that could mimic the derived slope in detail in Section~\ref{sec_causes}, we discuss our results in Section~\ref{sec_discussion}, and we finalize with some concluding remarks in Section~\ref{sec_conclusion}.

\section{Data log and data preparation}
\label{sec_Data_Reduction}

For this work we used archival data provided in the Mikulski Archive for Space Telescopes (Proposal ID 14767). Two transits of WASP-76~b were observed during the 16$^{th}$ of November, 2016, and the 17$^{th}$ of January, 2017. These data sets were collected using the HST STIS G430L grating. A third transit was observed on the 19$^{th}$ of February, 2017, but this time with the G750L grating. The G430L data set consists of 164 spectra spanning the two transits covering the wavelength range between 2900 and 5700~\AA. The G750L data set consists of 81 spectra covering the wavelength range between 5250 and 10300~\AA. Each transit spans five spacecraft orbits, and the visits are such that the third and fourth spacecraft orbits contain the center of the transit, which provides good coverage between second and third contact, as well as an out-of-transit baseline before and after the transit. The exposure time is fixed to 140 seconds in all cases, and the timestamp of each exposure is converted to Barycentric Julian Dates using the mid-time of the exposure and \cite{Eastman2010}'s tool.

During the reduction process we cleaned the data from cosmic ray hits using the \textit{cosmicrays} task in IRAF, which locates and removes cosmic rays using statistical modelling \citep{wells1994}. As this IRAF task is usually used and thought to work on images of stars rather than spectral images, we carried out a detailed process of cosmic ray extraction. Here, we divided one of our spectral images into three images, one image containing most of the light along the spectral trace, and two images containing the background, above and below the trace. The two images containing the background should only have light from cosmic rays, and we can thus identify the pixels contaminated from cosmic rays without confusing them with pixels containing light from the star. We used these images to train the selection criteria in the \textit{cosmicrays} task, as the input parameters are fundamental for the success of the extraction.

To determine the spectral trace we used IRAF's task  \textit{APALL}, with pixel-dependent polynomials which order ranged from first to tenth. An eight order polynomial was finally chosen, as this minimized the residuals of the fitted trace. Following \cite{wasp33b}, we used the background subtraction that comes with the \textit{APALL} task. The function for removing the background was set to second order and the predefined background regions were set to be as far away from the spectral trace as possible ($\sim$ 50 pixels/20 full width at half maximum, FWHM) while still having a substantial width containing purely background, of $\sim$10 pixels at each side of the spectral trace. Following the example of \cite{huitson} and \cite{wasp33b}, the aperture extraction was performed using \textit{APALL}. First, we extracted fluxes using several apertures, specifically from 10 to 50 pixels in steps of 1 pixel. Then, we produced white transit light curves (this is, light curves produced integrating fluxes in all wavelengths) and we computed the standard deviation of the off-transit data points. As final aperture we chose the one minimizing this standard deviation. The final apertures were fixed to 19 pixels for the G430L grating, and 43 pixels for the G750L grating. Our final aperture sizes include WASP-76's companion \citep[see e.g.,][]{Wolltert2015,Ngo2016}. The adequate treatment for third light contamination that we carry out to correctly derive the transmission spectrum of WASP-76~b is detailed in Section~\ref{sec:TLC}.

The STIS spectra were used to create both a white light photometric time series, where the flux is integrated over all wavelengths for each exposure, and chromatic light curves, where custom wavelength bands were chosen, integrating the flux in each wavelength bin for each exposure. The resulting light curves exhibit all the expected instrumental effects other authors have also encountered \citep{sing11,sing15,huitson,nikolov14,nikolov15}, though first described by \cite{brown2001}, and we follow their example to handle them. The main instrument related systematic effect is due to the thermal breathing of HST, which warms and cools the telescope during its $\sim$96 min day/night cycle \citep{sing13}. This results in changes in the point spread function and in the central position of the spectrum. This was accounted for by fitting a fourth order polynomial phased to the HST period to the fluxes, in simultaneous to the transit model. As carried out by other authors, we also included in our detrending model a linear slope in time. Although this was our final detrending model setup, we also tried a second and third order polynomial phased to the HST period, a second order polynomial as a function of time, and a linear combination between the time-dependent spatial and wavelength shifts of the spectra (see Section~\ref{sec:detrending_model}), combining them in all possible ways. Our final detrending choice is favoured by the minimization of the Bayesian Information Criterion, \mbox{BIC = $\chi^2$ + k ln N}, where k is the number of model parameters, $\chi^2$ is computed between data and sub-models, and N is the number of data points.

When the telescope is moved to a new pointing position it takes approximately one spacecraft orbit to thermally relax, which compromises the stability of the first orbit \citep{huitson}. In our first analysis we tried to keep the first orbit, but found that our detrending was not satisfactory, as a fourth order polynomial could not remove the large systematic trends. Therefore, we did not consider the first orbit of each transit during our analysis. For STIS it is also known that the first exposure of each spacecraft orbit is significantly fainter than the remaining exposures. To bypass this problem, \cite{sing11,sing15,huitson,nikolov14,nikolov15} have set the exposure time of the first exposure to only 1 second. The main idea is to discard this exposure without suffering from a significant loss of valuable observing time. The observations analyzed in this work were obtained using this observing strategy.

For our transit light curves we computed the individual spectro-photometric errors following \cite{vonessen2017}. For this end, we used the formalism provided by IRAF's photometric errors, i.e.,

\begin{equation}
    \epsilon^2 = \frac{F/g + A \sigma^2 + (A^2 \sigma^2)/N}{F},
\end{equation}

\noindent where F is the integrated flux inside the wavelength band and aperture, A is the area inside the band and aperture, $\sigma$ is the standard deviation of the background, N the number of pixels in the background, and g is the gain of the detector. As also pointed out by \cite{vonessen2017}, errors produced in this fashion are often underestimated, as they follow a photon-noise-only distribution, which is most likely unrealistic. We therefore scaled our errors with the standard deviation of the residual light curves. These were obtained subtracting to the fluxes a first model obtained from a quick least-squared fit. 

\section{Data modelling and fitting parameters}
\label{sec_dat_mod_fit_param}

\subsection{Transit model and limb darkening coefficients}
\label{sec:TM_LD}

The light curves were modelled using \cite{mandel}. Here, the model parameters are the orbital inclination, i, the mid-transit time, T$_0$, the orbital period, P, the semi-major axis in stellar radii, $\mathrm{a/R_s}$, the planet-to-star radius ratio, $\mathrm{R_P/R_S}$, the third light contribution, described as the flux ratio between the stellar companion and the main star, and the limb darkening coefficients corresponding to a quadratic limb-darkening law: 

\begin{equation}
\frac{I(\mu)}{I(1)} = 1 - a(1-\mu) - b(1-\mu)^2,
\label{eq_qua}
\end{equation}

\noindent where $I(1)$ is the specific intensity at the centre of the
stellar disk, $a$ and $b$ are the linear and quadratic limb darkening
coefficients (LDCs), respectively, and $\mu = \cos(\gamma)$, where
$\gamma$ is the angle between the line of sight and the emergent
intensity. We chose the simplest law as the differences in transit
shape between a quadratic and a four-parametric non-linear limb
darkening law are beyond the precision of our data \citep[see
  e.g,][for a similar choice]{Sotzen2020}. Nonetheless, we would like
to emphasize that the precision given by limb-darkening coefficients
only reflects the precision in the fit between stellar intensity
models and the limb darkening model \citep{vonessen2017}, regardless
of the choice of law. It does not reflect the real accuracy at which
we know the radial profiles of stellar intensities \citep[see
  e.g.,][]{White2013,Boyajian2015,Kervella2017}, reason why we
disbelieve in applying changes that might have an impact in the third
or fourth decimal of the limb-darkening coefficients, as these would
not be supported by our knowledge on stellar physics. To calculate the
customized LDCs for each wavelength bin, we used angle-dependent,
specific intensity spectra from PHOENIX \citep{phoenix} with main
stellar parameters corresponding to the effective temperature,
\mbox{$\mathrm{T_{eff}}$ = 6200} K, the surface gravity, \mbox{log(g)
  = 4.5}, and the metallicity, \mbox{[Fe/H] = 0.00}, closely matching
the values of WASP-76, which are \mbox{$\mathrm{T_{eff}}$ = 6250 K},
\mbox{log(g) = 4.4} \citep{seidel2019} and \mbox{[Fe/H] = 0.19}
\citep{star}. As performed by \cite{vonessen2017} and
\cite{claret2011}, to compute the limb darkening coefficients we
neglect the data points at small $\mu$'s, specifically those between
\mbox{$\mu$ = 0} and \mbox{$\mu$ = 0.064}. The uncertainties of the
linear and quadratic LDCs are obtained from $\chi^2$ maps,
specifically choosing the values of the LDCs at which $\Delta\chi^2$ =
1. All the LDCs used for the chromatic light curves are presented in
Table \ref{tab_bins}.

\begin{figure}[ht!]
	\centering
	\includegraphics[width=0.49\textwidth]{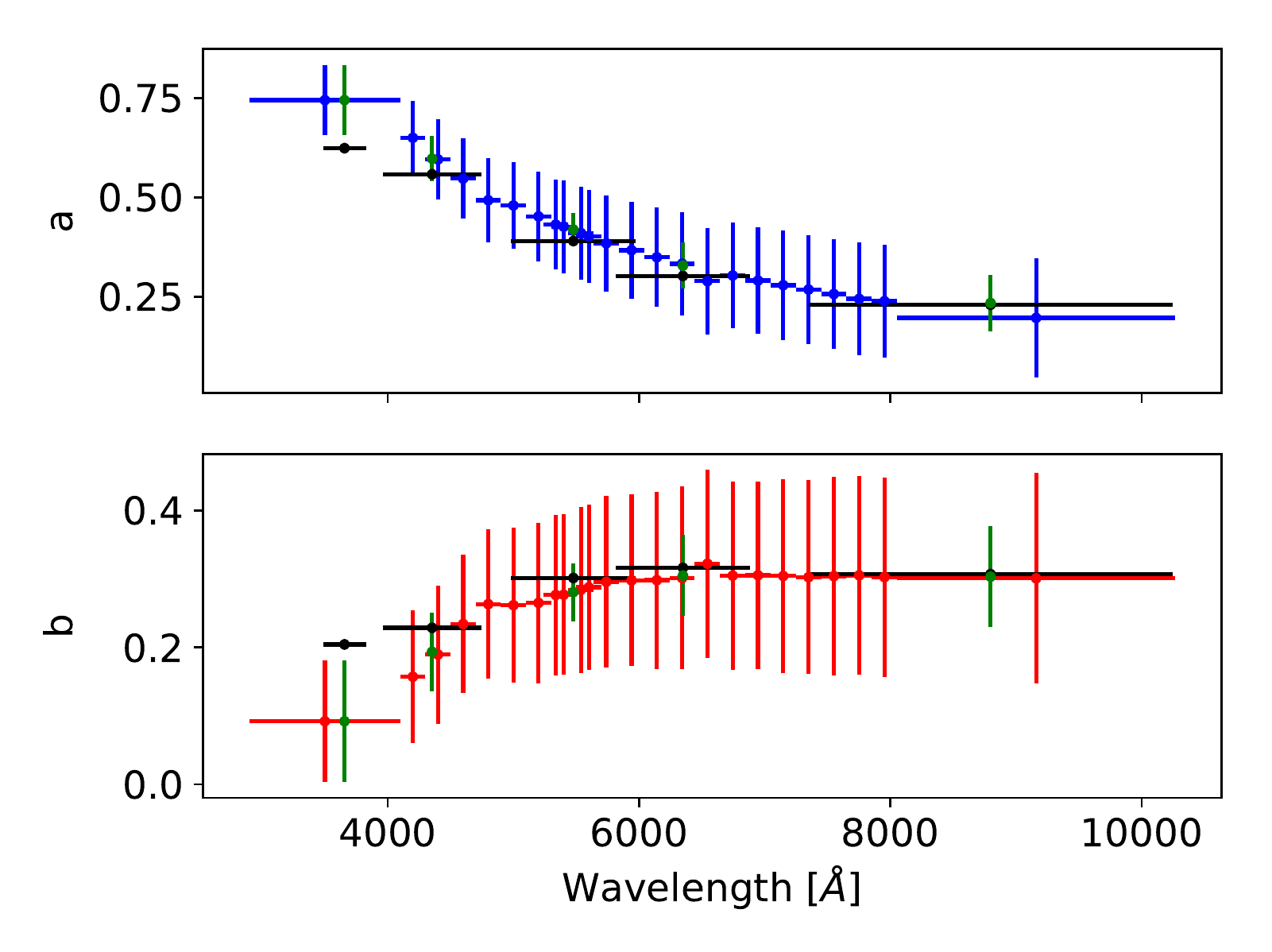}
	\caption{Computed limb darkening coefficients (LDCs) for the chromatic light curves in red (linear LDC) and blue (quadratic LDC), compared to the LDCs for the $UBVRI$ filters taken from \cite{claret2011} in black. In all cases, horizontal lines correspond to the width of the integration band for the custom wavelength bins, and the FWHM for the broad band filters. The green dots represent average values of the LDCs computed in this work, coinciding with the wavelength coverage of each of the $UBVRI$ filters. The vertical lines show their uncertainties.}
	\label{fig_ab}
\end{figure}

To assess the quality of our procedure, we compared our LDCs with the ones computed by \cite{claret2000} and \cite{claret2011} for the Johnson-Cousins $U$, $B$, $V$, $R$ and $I$ filters. A comparison between LDCs can be seen in Figure~\ref{fig_ab}. In the figure, the blue and red data points are our calculated LDCs for each wavelength bin for the linear, $a$, and quadratic, $b$, coefficients specified in Equation~\ref{eq_qua}. The horizontal lines represent the width of the bins, and the vertical lines show the uncertainty on each LDC. To compare our costumed LDCs with those computed by \cite{claret2000} and \cite{claret2011} for the $UBVRI$ filters, we compute averages and errors of those LDCs contained within the FWHM of each broad band filter. The derived values are shown in the Figure in green. The best match between our customized LDCs and published ones comes from the values reported by \cite{claret2011}, shown in black in Figure~\ref{fig_ab}. The resulting values are presented in Table \ref{tab_UBVRI}. \cite{claret2011} does not provide uncertainties. In consequence, these are absent in the Table and the Figure. As Figure~\ref{fig_ab} reveals, ours and \cite{claret2011}'s LDCs match at all wavelengths considering 1-$\sigma$ uncertainties. It is worth to mention that the errors provided in this work are merely statistical, and do not realistically reflect the accuracy at which we know any limb-darkening coefficients.

\begin{table*}[]
	\centering
	\captionsetup{justification=centering,margin=1.5cm}
	\caption{LDCs for the $U$, $B$, $V$, $R$ and $I$ filters obtained from \cite{claret2011} (C11), and computed in this work (TW). The coefficients, a and b, correspond to those in equation~\ref{eq_qua}.}
	\begin{tabular}{llllll}
		\hline
		& $U$	         & $B$       	& $V$     		& $R$      		& $I$      		\\ \hline
		$a_\mathrm{C11}$ & 0.6234  		& 0.5578  		& 0.3898 		& 0.3021 		& 0.2296 		\\ 
		$a_\mathrm{TW}$  & 0.7442 $\pm$ 0.0889 & 0.5971 $\pm$ 0.1704	& 0.4188 $\pm$ 0.3287 & 0.3280 $\pm$ 0.2875	& 0.2340 $\pm$ 0.2850 \\
		$b_\mathrm{C11}$ & 0.2043  		& 0.2286  		& 0.3013 		& 0.3162 		& 0.3069 		\\ 
		$b_\mathrm{TW}$  & 0.09210 $\pm$ 0.0889 & 0.1936 $\pm$ 0.1727 & 0.2807 $\pm$ 0.3387 & 0.3048 $\pm$ 0.2965 & 0.3031 $\pm$ 0.2950 \\ \hline
	\end{tabular}
	\label{tab_UBVRI}
\end{table*}

\subsection{Detrending model}
\label{sec:detrending_model}

HST light curves show non-Gaussian, correlated noise (see Section~\ref{sec_Data_Reduction}). In consequence, as previously done \citep{sing11,sing15,huitson,nikolov14,nikolov15}, we apply orbit-to-orbit flux corrections by fitting a fourth degree polynomial phased to the orbit of HST (henceforth, model M4) to account for the thermal breathing of HST, and a linear slope (henceforth, model M1) in time. The resulting detrending function is described as follows:

\begin{equation}
\label{eq:detrending}
\begin{split}
    f(t) = & ~(a_0 + a_1 \cdot x + a_2 \cdot x^2 + a_3 \cdot x^3 + a_4 \cdot x^4) \\
    &\cdot (c_0 + c_1 \cdot (t - T_0)),
\end{split}
\end{equation}

\noindent where $a_0, a_1, a_2, a_3$ and $a_4$ are the detrending coefficients for the fourth degree polynomial, and $c_0$ and $c_1$ are the detrending coefficients for the linear slope. In addition, $t$ is the time, $T_0$ is the mid-time of the transits, and:

\begin{equation}
    x_i = \frac{t_i - t_0}{\mathrm{per_{HST}}} - \mathrm{int}\left(\frac{t_i - t_0}{\mathrm{per_{HST}}}\right).
\end{equation}

\noindent Here, $t_0$ corresponds to the first time stamp for each transit, and per$_\mathrm{HST}$ corresponds to the period of HST's orbit. These functions are fitted simultaneously to the transit model.

Earlier studies using HST data \citep[see e.g.,][]{sing11,sing13,huitson} have also included systematic trends which correlate with the X and Y detector positions of the spectra. To analyze if this was necessary, we computed both X (model X, MX) and Y (MY) shifts. First, we determined the shift in the Y detector position by fitting multiple Gaussian functions to the stellar spectrum in its spatial direction. The reference position of the trace per X pixel is represented by the mean of the Gaussian function. For this exercise we considered the first spectral exposure of each transit to be the reference spectrum, and found the shift in the spectral trace by comparing the other spectra to this, making one-to-one differences of these mean values, that were afterwards averaged. Then, we computed the shift in X detector position by selecting 3 deep absorption lines from the stellar spectra. To each one of these lines we fitted a Gaussian function, and used their means as central wavelength positions. Equivalently to the Y shifts, we chose the first frame as the reference one. Computing one-by-one differences between the line centers of the reference frame and the subsequent spectral exposures, and averaging these differences in time, we determined the X shifts. To quantify to which extent were the chromatic light curves affected by the X and Y detector positions of the spectral traces, we made use of the Pearson correlation coefficient (PCC), computed between our residuals and the X and Y trends determined as just explained. After finding no strong correlation between these pairs (PCC$<$0.2 in all cases), we decided not to include these components in the detrending model. This choice was also supported making use of the BIC. In Table~\ref{tab:BIC_vals} we list some of the most relevant detrending models and their corresponding BIC values, obtained averaging the BIC's computed from each one of the three white light curves. Besides the BIC minimization, in this work we have considered $\Delta$BIC $<$ -5 between two given models as strong evidence to which model is more likely \citep{Kass1995}.

\begin{table}[ht!]
  \caption{\label{tab:BIC_vals} BIC minimization computed considering different detrending models. From left to right the detrending model, the number of detrending parameters fitted in each case, NDP, and the Bayesian Information Criterion (BIC).} 
  \centering
  \begin{tabular}{lc c }
    \hline \hline
    Model                                 &  NDP  &     BIC   \\
    \hline
    (1) M1                                &  2    &  178.5 \\
    (2) M4                                &  5    &  164.2 \\
    (3) M1 $\times$ M4                    &  7    &  137.9 \\
    (4) M1 $\times$ M4 $\times$ MX        &  8    &  142.3 \\
    (5) M1 $\times$ M4 $\times$ MY        &  8    &  143.1 \\
    (6) M1 $\times$ M4 $\times$ (MX + MY) &  9    &  143.7 \\
    \hline
  \end{tabular}
\end{table}

\subsection{Generalities on our fitting procedure}

As carried out by \cite{wasp33b}, to derive the model parameters we fitted all the light curves simultaneously using Markov-chain Monte Carlo (MCMC), all wrapped up in PyAstronomy\footnote{www.hs.uni-hamburg.de/DE/Ins/Per/Czesla/PyA/PyA/index.html} \citep{Patil2010,Jones2001}. For both white (flux integrated in all wavelengths) and chromatic (flux integrated in narrow wavelength bands) light curves, we carried out our parameter fitting procedure in two stages. The first stage was carried out specifically to quantify the amount of correlated noise in our light curves, in order to determine reliable spectro-photometric error bars. The second stage was carried out to determine reliable error bars for the fitting parameters, using enlarged photometric error bars that accounted for correlated noise.

In general, the iterations used for the white light curves are \mbox{200\ 000} and \mbox{1\ 000\ 000} for the first and second stages, respectively, with a burn in of the initial 20\% samples. Equivalently, for the chromatic light curves we iterated \mbox{400\ 000} and \mbox{500\ 000} times, with a burn-in of the initial \mbox{100\ 000} samples. From the posterior distributions we computed the mean and standard deviation \mbox{(1-$\sigma$)}, and used them as our best-fit values and uncertainties, respectively. The MCMC chains were checked for convergency by visual inspection, which in turn was used to set the burn-in. We also divided the chains in three sub-chains, and computed from each one of them the usual statistics. We considered a chain to converge if the derived parameters were consistent between each other at 1-$\sigma$ level.

Finally, we chose the starting values for our MCMC chains during the first stage to be the ones specified in Table \ref{tab_ait0}, adopted from \cite{seidel2019}. The uncertainty in the Gaussian priors were set to be three times the author's uncertainties for each parameter. For the second stage we used the best-fit values from the first run, considering uncertainties at a 3-$\sigma$ level.

\subsection{First MCMC run: correlated noise}
\label{sec_fit_strag_corr_noise}

To compute the amount of correlated noise in our light curves, we followed the methods of \cite{corrnoise}. To do so, we computed residuals from our first MCMC fit, subtracting the transit-times-detrending model from our transit light curves. As each HST orbit takes about 45 minutes and the ingress/egress duration of WASP-76~b is approximately half of that, we divided each HST orbit into 2 bins of equal duration (M = 2$\times$4 orbits = 8), and calculated the number of points in each bin, N. If the data are not affected by correlated noise, they should follow the expectation of independent random numbers, 

\begin{equation}
    \sigma_N = \sigma_1 N^{-1/2} [M/(M-1)]^{1/2},
\end{equation}

\noindent where $\sigma_1$ is the variance of the unbinned data, and $\sigma_N$ is the variance of the binned data, with the following expression:

\begin{align}
    \sigma_N = \sqrt{\frac{1}{M} \sum^M_{i=1}( \langle \hat{\mu}_i\rangle - \hat{\mu}_i)^2}.
\end{align}

\noindent Here, $\hat{\mu}_i$ is the mean value of the residuals per bin, and $\langle \hat{\mu}_i\rangle$ is the mean value of all the means. If the data are affected by correlated noise, each $\sigma_N$ value would differ by a factor of $\beta_N$ from their expectation value. $\beta$ is an estimation of the strength of the correlated noise computed from averaging certain $\beta_N$'s, so a $\beta = 1$ means no correlated noise. As we chose to divide each HST orbit only in two, $\beta_N = \beta$. We finally increased the size of the error bars by this factor. To place our results in context with other work, \cite{nikolov14} find $\beta$ values of 1.2 and 1.3 in their white light curves for observations carried out with the G430L grating, and a $\beta$ of 1 for observations with the G750L grating. \cite{evans2018} find $\beta$ values of 1.29, 1.16, and 1.36 for the same gratings. Our computed $\beta$ values are 1.35, 1.30 and 1.55, respectively. 

\subsection{Second MCMC run: best-fit parameters and uncertainties}

For all the light curves the central transit time, T$_0$, the semi-major axis in units of stellar radius, a/$\mathrm{R_s}$, the orbital inclination, i, the planet-to-star radius ratio, $\mathrm{R_p/R_s}$, and the detrending parameters specified in Equation~\ref{eq:detrending} were fitted simultaneously. As T$_0$, a/$\mathrm{R_s}$, and i are values that are wavelength-independent, we treated them as equal for all the white and chromatic light curves. In other words, we only fitted one of each parameter for all the wavelength bins and the three light curves combined. Contrary to this, the $\mathrm{R_p/R_s}$ parameters were treated as equal in each wavelength bin for the two transits in the G430L grating, and differently in different wavelength bins. All the detrending parameters and $\mathrm{R_p/R_s}$ have uniform priors. The transit parameters T$_0$, a/$\mathrm{R_s}$, and i have Gaussian priors, as we know these from \cite{seidel2019}. We also use the value from \cite{seidel2019} for the period, but we considered the orbital period as fixed, as it is known with a very high degree of precision and, in consequence, will have a negligible impact in our light curve fitting. 

\section{Results}
\label{sec_results}

\subsection{Third light of another star in the aperture}
\label{sec:TLC}

If there is a companion whose light falls into the photometric aperture, its third light contamination modifies the transit depth of the planet. When the companion is of later spectral type than the planet host, the third light contribution becomes stronger toward longer wavelengths. That means, the dilution of the transit depth strengthens towards redder wavelengths, the derived apparent planet-star radius becomes smaller, thus it might mimic a scattering slope in the transmission spectrum \citep{Southworth2016,Mallonn2016}. In the case of WASP-76, there is a well characterized companion with a separation of $\sim$0.4 arcseconds. Considering STIS plate scale, this translates into approximately 8 pixels. Some literature values of its magnitude contrast relative to WASP-76, coinciding with STIS G750L wavelength range, are \mbox{2.58 $\pm$ 0.27 mag} in the SDSS i' band \citep{Ginski2016}, and \mbox{2.51 $\pm$ 0.25 mag} and \mbox{2.85 $\pm$ 0.33 mag} in the SDSS i' and z' band, respectively \citep{Wolltert2015}. 

\begin{figure}[ht!]
    \centering
    \includegraphics[width=.5\textwidth]{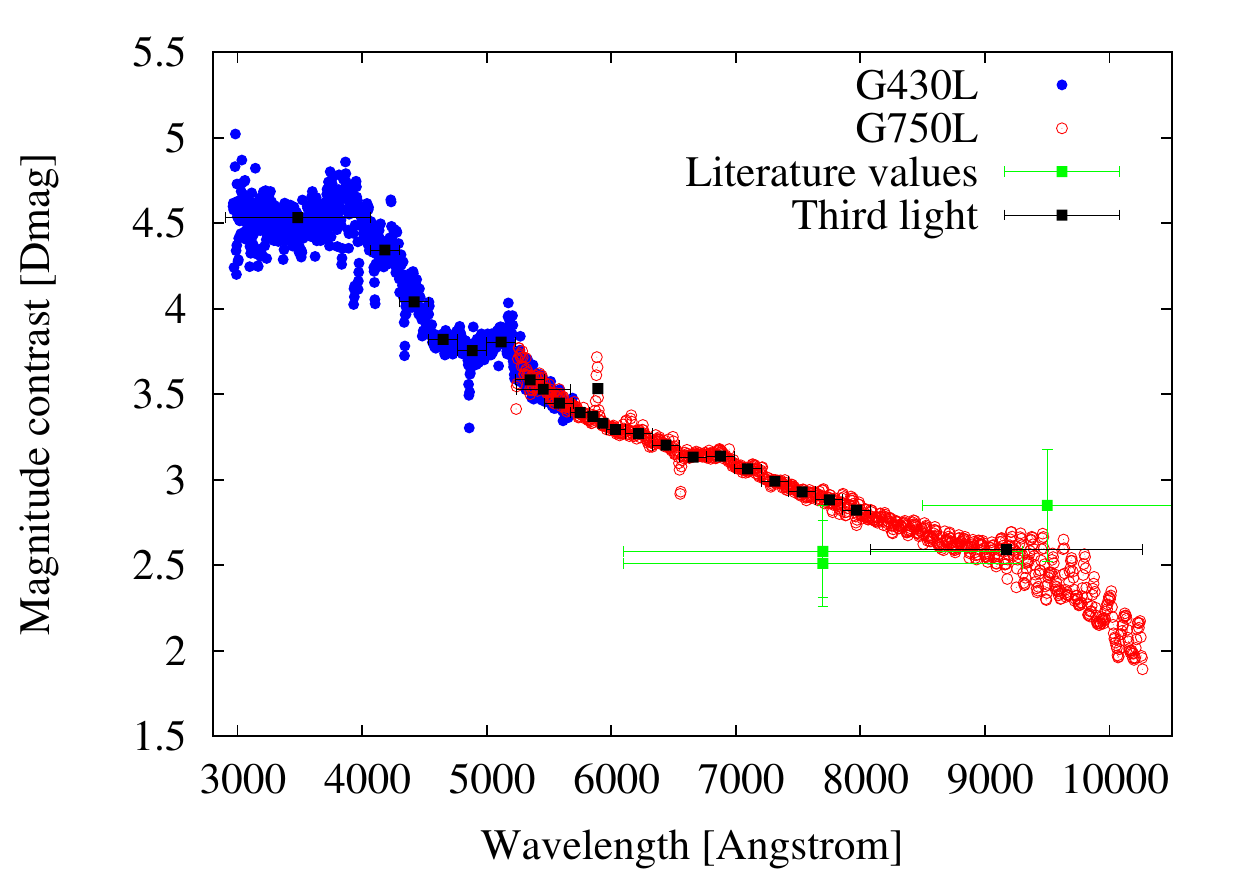}
    \caption{\label{fig:companion} Third light contribution of WASP-76's companion in $\Delta$mag as a function of wavelength. Blue filled circles correspond to the values derived from the G430L grating, while red empty circles belong to the G750L grating. In green squares we show literature values found in \cite{Wolltert2015} and \cite{Ginski2016}. Uncertainties for the magnitudes are given at 1-$\sigma$ level. In black squares we show our flux-weighted third light contribution. Horizontal bars show either the filter FWHM or the width of the integration band.}
\end{figure}

Either while computing the transit depth from the white light curves (Section~\ref{sec:WLC}) or from the chromatic transits in Section~\ref{sec:TS}, the third light treatment to the transit light curves is the same, and is described as follows. From the spectroscopic 2D images we can disentangle the two PSFs of WASP-76 and its companion, so computing the third light contribution directly from the spectra is straight forward to do. To compute the flux ratio between companion and star we proceeded similarly to \cite{Mallonn2016}. Per wavelength element we produced a cut in the spatial direction. To this profile, we fitted the sum of two Gaussian functions with same standard deviation using least-square minimization. The final third light contribution is nothing more than the ratio between the integrated fluxes of each star, weighted by each Gaussian profile. To increase the signal-to-noise of our wavelength-dependent third light contribution, we reproduced this process over each one of the available spectra. Finally, we averaged in the wavelength direction. Figure~\ref{fig:companion} shows our derived third light contribution in $\Delta$mag, compared to literature values \citep{Wolltert2015,Ginski2016}. Even though our computation was performed independently per grating and image, there is a perfect overlap between the two gratings, and between the data sets of the two transits in the G430L grating. As the figure reveals, the magnitude contrast quickly increases with the increase in wavelength, which is typical of a companion of late spectral type. 

From the derived wavelength-dependent contamination, for a specific chromatic light curve we computed the third light contribution from the weighted mean within each wavelength bin, using as weight the flux of WASP-76 alone, which in turn was also obtained averaging the fluxes of all available spectra. In this way, values where WASP-76's flux is larger, have a higher weight. The third light contribution considered in our transit fitting is plotted in Figure~\ref{fig:companion} with black squares.

\subsection{White light curve}
\label{sec:WLC}

\begin{figure*}[ht!]
	\centering
	\includegraphics[width=0.49\textwidth]{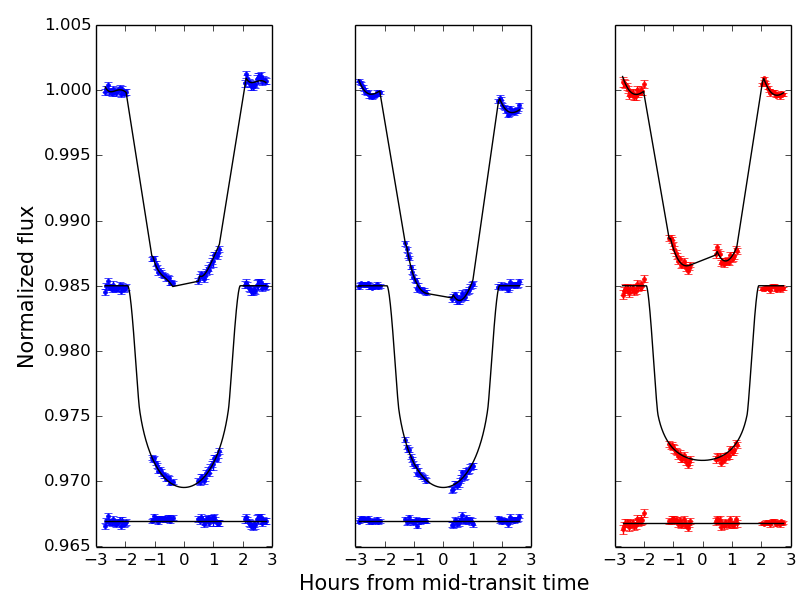}\hfill%
	\includegraphics[width=0.49\textwidth]{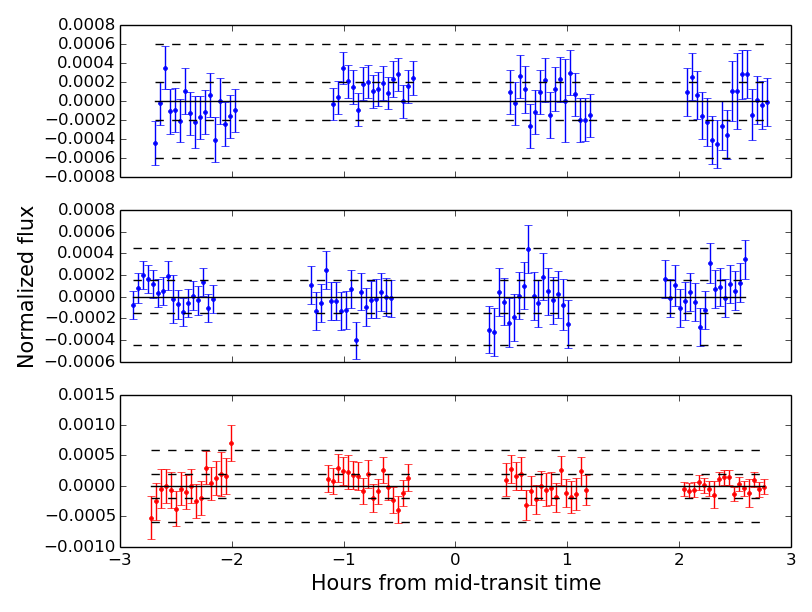}
	\caption{HST white light curves for the three transits, shifted to their individual mid-transit times and shown in hours. Blue transits correspond to those taken with the G430L grating, and red to the one collected with the G750L grating. {\it Left:} from top to bottom we show the transit light curves with the systematic effects, along with our best-fit combined model (transit times detrending), the detrended data along with our best-fit transit model, and the residual light curves after both components were subtracted. {\it Right:} zoom in to the residuals. Dashed horizontal lines indicate $\pm$1 and $\pm$3 times the standard deviation of the residuals to guide the eyes.}
	\label{fig_white}
\end{figure*}

For our white light curves we integrated the light over all wavelengths for each exposure. For the G430L grating the specific wavelength range is from 2900 to 5700~\AA, and for the G750L grating the wavelength range is from 5250 to 10300~\AA. Our white light curves are presented in Figure~\ref{fig_white}, {\it left}, both containing the raw photometry with the systematic noise on the top, the detrended light curves in the middle, and the residuals in the bottom. A zoom in to the residuals is shown in Fig.~\ref{fig_white}, {\it right}. In all cases, with continuous black lines we show our best-fit combined model on top (transit times detrending), and our best-fit transit model only in the middle of the figure. In addition, Table~\ref{tab_ait0} shows our derived values for the transit parameters that were fitted in this work. For comparison, the values from \cite{west2016} and \cite{seidel2019} are also shown. Their reported $\mathrm{R_p/R_s}$'s are computed considering the spectral bands 3872 - 6943~\AA, 3900 - 6800~\AA, and 5850.24 - 5916.17~\AA, respectively. As the table reveals, our derived transit parameters, a/R$_s$ and i, are consistent with previous work at 1-$\sigma$ level. Posterior distributions for some of the fitted parameters can be seen in Appendix~\ref{Appendix}. To investigate the impact of our choice of detrending over the transit depth, Fig.~\ref{fig:triangle1} shows the correlations between the transit parameters and the coefficients for the linear slope, c$_0$ and c$_1$, Fig.~\ref{fig:triangle2} those between the transit parameters and the coefficients for the fourth degree polynomial, a$_0$, a$_1$, a$_2$, a$_3$ and a$_4$. Besides the very well documented correlation between semi-major axis and inclination seen in both figures, we computed the Pearson's correlation coefficient:

\begin{equation}
  r_{xy} = \frac{\sum_{i=1}^n\ (x_i - \mu_x) (y_i - \mu_i)}{[\sum_{i=1}^n\ (x_i - \mu_x)^2 \sum_{i=1}^n\ (y_i - \mu_y)^2]^{1/2}}\,,
\end{equation}

\noindent for $x$ the planet-to-star radius ratio and $y$ the
different detrending parameters. For the linear slope case, the
r$_{xy}$ coefficients were found to be between -0.03 and 0.05, and for
the fourth degree polynomial, between -0.05 and 0.04. As both cases
reflect a low correlation, we did not investigate the impact of the
detrending coefficients any further.

\begin{table*}[ht!]
\centering
\captionsetup{justification=centering,margin=1.5cm}
\caption{Computed transit parameter values from the white light curves
  (WLC) compared to those from \cite{seidel2019} and
  \cite{west2016}. The wavelengths used to compute the transit light
  curves are different, so the corresponding $\mathrm{R_p/R_s}$'s are
  not to be compared. $\Delta\lambda$ indicates the wavelength range
  in which our white light curves were integrated.}
\begin{tabular}{cccc}
\hline\hline
                    & This work (WLC)               &  \cite{seidel2019}         & \cite{west2016}          \\ 
\hline
$\mathrm{R_p/R_s}$  & G430L: 0.11122 $\pm$ 0.00032  & 0.10824 $\pm$ 0.00081      & 0.1090 $\pm$ 0.0007  \\
                    & G750L: 0.11026 $\pm$ 0.00029  &                            &                          \\
a/R$_S$             & 4.036 $\pm$ 0.032             & 4.08$^{+0.02}_{-0.11}$     & 4.102 $\pm$ 0.062        \\
i ($^{\circ}$)      & 88.21 $\pm$ 0.95              & 86.72$^{+1.72}_{-1.18}$    & 88.0$^{+1.3}_{-1.6}$     \\
T$_0$ (BJD$_{TDB}$) & 7709.59863 $\pm$ 0.00012      & 8080.62487 $\pm$ 0.00018   & 6107.85507 $\pm$ 0.00034 \\ 
$\Delta\lambda$ (\AA) &  G430L: [2900-5700] - & & \\
                      &  G750L: [5250-10300] & & \\
\hline
\end{tabular}
\label{tab_ait0}
\end{table*}

\subsection{Transmission spectrum of WASP-76~b}
\label{sec:TS}

\begin{figure*}[ht!]
	\centering
	\includegraphics[width=1.0\textwidth]{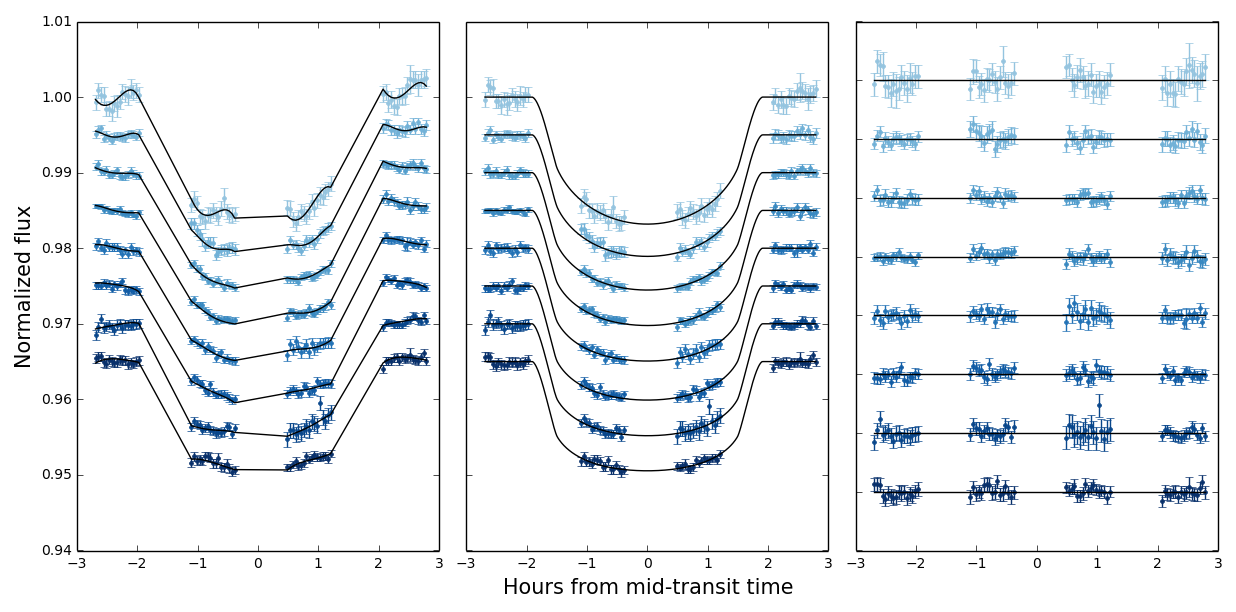}
	\caption{Spectral bin transit light curves obtained with the G430L grating on HST/STIS. The light curves have an arbitrary offset in relative flux and are ordered by wavelength (see Table~\ref{tab_bins}). Here, the bin at shortest wavelengths is located at the top (lightest in color in the figure) and the bin at the longest wavelengths is located at the bottom (darkest in color in the figure). The individual photometric errors are scaled with the standard deviation of the residuals and further increased by $\beta$, as described in Section~\ref{sec_fit_strag_corr_noise}. {\it Left:} Light curves overplotted with the best-fit combined model. {\it Middle:} Light curves corrected for systematic effects, with the best-fit transit model overplotted. {\it Right:} Residuals.}
	\label{fig_trans1}
\end{figure*}

\begin{figure*}[ht!]
	\centering
	\captionsetup{justification=centering}
	\includegraphics[width=1.0\textwidth]{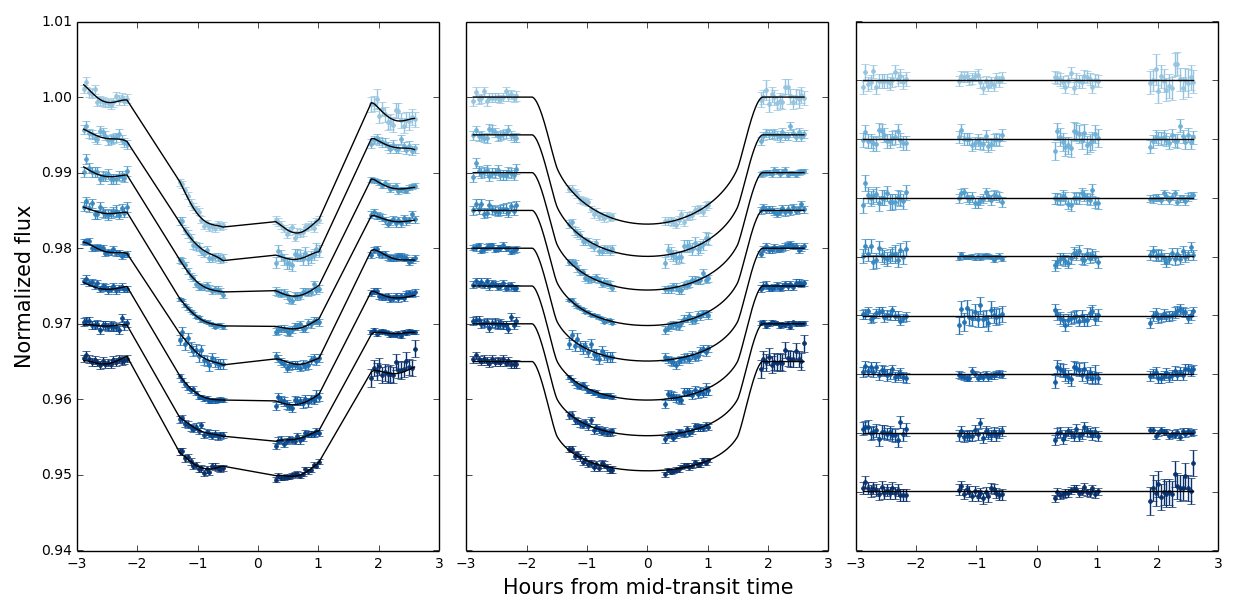}
	\caption{Same as Figure \ref{fig_trans1}, but for the second visit.}
	\label{fig_trans2}
\end{figure*}

\begin{figure*}[ht!]
	\centering
	\captionsetup{justification=centering}
	\includegraphics[width=1.0\textwidth]{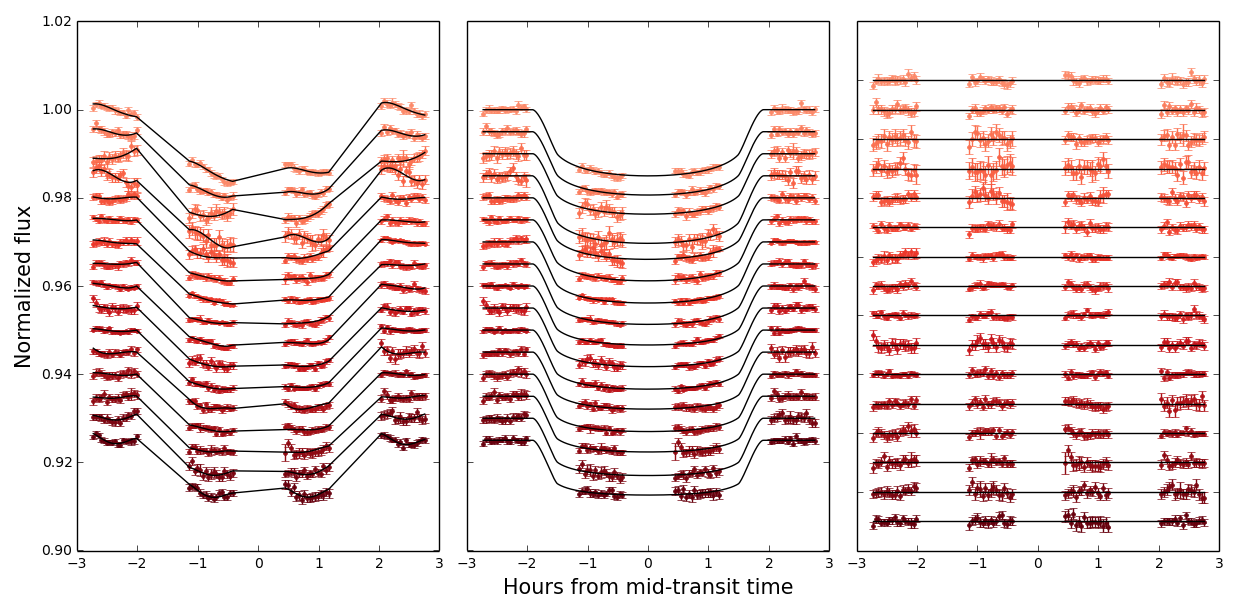}
	\caption{Same as Figure \ref{fig_trans1}, but for the G750L spectral bins.}
	\label{fig_trans3}
\end{figure*}

\begin{figure*}[ht!]
	\centering
	\includegraphics[width=\textwidth]{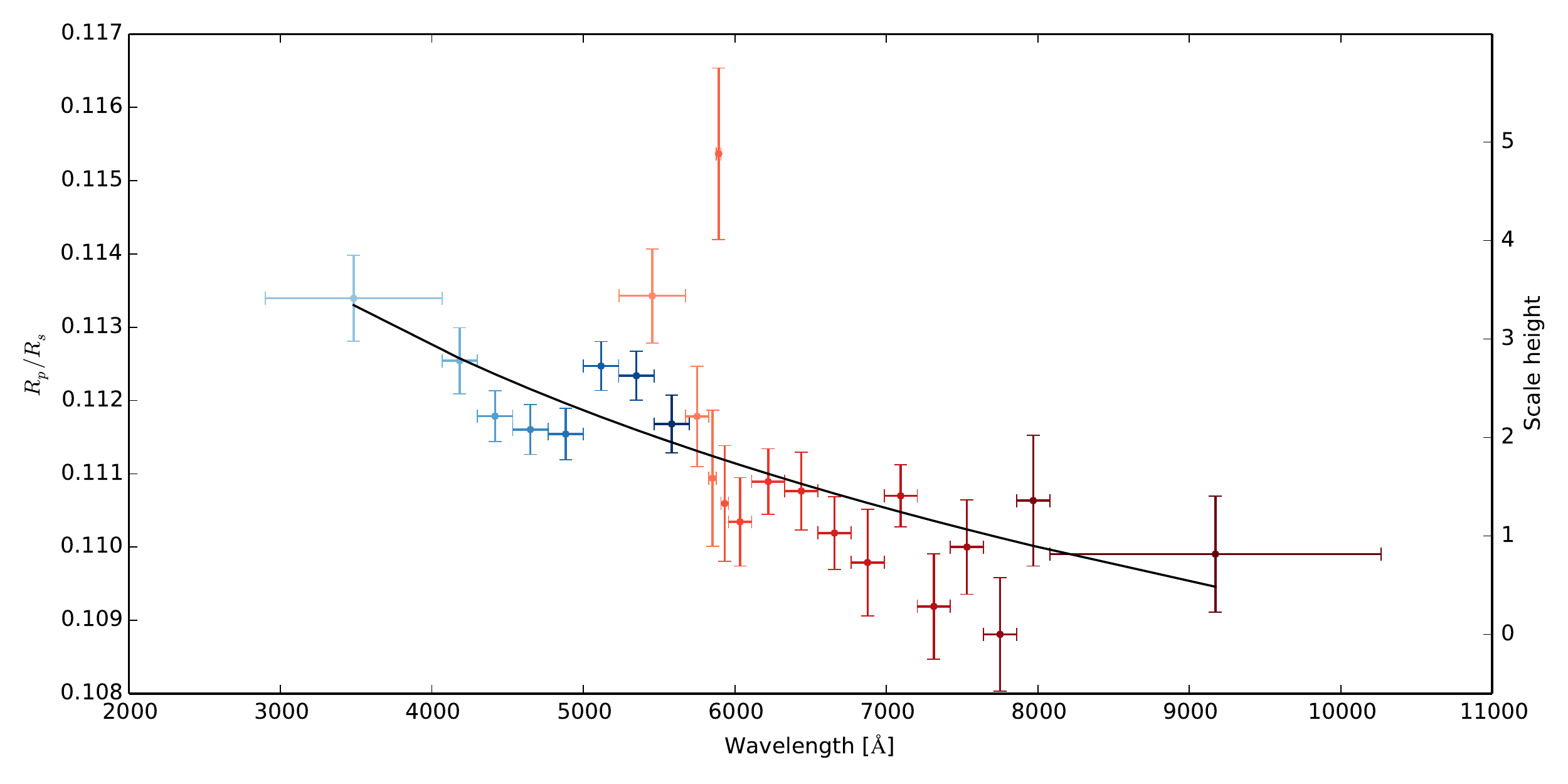}
	\caption{Broad-band transmission spectrum for WASP-76~b. Blue dots correspond to the G430L grating, while red dots correspond to the G750L grating. The colors match those in Figures~\ref{fig_trans1}, \ref{fig_trans2} and \ref{fig_trans3}. The black line shows the fitted Rayleigh slope from Equation~\ref{eq_AD}. The left vertical axis shows the planet-to-star radii ratio, while the right vertical axis shows the same variability in scale heights.}
	\label{fig_transmission}
\end{figure*}

Initially, the spectra were divided into eight and thirteen wavelength
bins for the blue and red gratings, respectively, and the
$\mathrm{R_P/R_S}$ were computed for each bin. After
\cite{seidel2019}'s work, we recomputed the transmission spectrum in
the same way, but choosing three narrow bands around sodium,
specifically 20~\AA\ wide, centered at 5892.9~\AA. The specific values
for the wavelength bins across the whole spectrum can be seen in
Table~\ref{tab_bins}, along with the scatter and LDCs used for each
bin. We also tested 20~\AA\ wide wavelength bins centered at the line
cores of the potassium doublet. However, no additional absorption
compared to the adjacent bands was detected. The chromatic light
curves, both before and after detrending, are plotted in Figure
\ref{fig_trans1}, Figure \ref{fig_trans2}, and Figure
\ref{fig_trans3}, along with their residuals. The broad-band
transmission spectrum can be seen in
Figure~\ref{fig_transmission}. The left vertical axis shows the
variability of the planet-to-star radii ratio, while the right
vertical axis shows the same variability but in scale
heights. Overplotted to the derived $\mathrm{R_P/R_S}$'s, we show our
best-fit Rayleigh slope (see Section~\ref{Rayleigh}) in continuous
black line. Similarly to the white light curves, to assess the impact
of detrending over the $\mathrm{R_P/R_S}$ we computed the Pearson's
correlation value between $\mathrm{R_P/R_S}$ and the detrending
coefficients. In all cases, r$_{xy}$ ranged between -0.21 and
0.18. The posterior distributions look similarly as
Figures~\ref{fig:triangle1} and \ref{fig:triangle2}. As the
correlation is in all cases so low, we do not investigate the impact
of detrending into the derived transmission spectrum, for being the
correlation between them negligible.

\begin{table*}[]
\caption{The wavelength bins, the $\mathrm{R_p/R_s}$ computed in each bin, their corresponding LDC, and the scatter of the resulting transit light curves in parts per million (ppm). The top 8 rows correspond to the G430L grating, and the remaining 13 correspond to the G750L grating. As there are two transits in the G430L grating, there are two values for the scatter.}
	\centering
	\captionsetup{justification=centering,margin=2cm}
	\begin{tabular}{ccccc}
		\hline
		Wavelength [\AA]  & Rp/Rs & a      & b          & Scatter (ppm)  \\ \hline
		G430L & & & & \\
		2900.00 - 4066.67 & 0.11339 $\pm$ 0.00059 & 0.749 $\pm$ 0.089 & 0.088 $\pm$ 0.089 & 729 / 455  \\ 
		4066.67 - 4300.00 & 0.11254 $\pm$ 0.00045 & 0.646 $\pm$ 0.093 & 0.162 $\pm$ 0.097 & 429 / 417 \\ 
		4300.00 - 4533.33 & 0.11178 $\pm$ 0.00034 & 0.59 $\pm$ 0.10 & 0.20 $\pm$ 0.10 & 304 / 319 \\ 
		4533.33 - 4766.67 & 0.11160 $\pm$ 0.00034 & 0.54 $\pm$ 0.10 & 0.24 $\pm$ 0.11 & 323 / 331 \\ 
		4766.67 - 5000.00 & 0.11154 $\pm$ 0.00035 & 0.48 $\pm$ 0.11 & 0.27 $\pm$ 0.11 & 362 / 342 \\ 
		5000.00 - 5233.33 & 0.11247 $\pm$ 0.00033 & 0.46 $\pm$ 0.11 & 0.26 $\pm$ 0.11 & 350 / 293 \\ 
		5233.33 - 5466.67 & 0.11233 $\pm$ 0.00033 & 0.43 $\pm$ 0.11 & 0.28 $\pm$ 0.12 & 461 / 291 \\ 
		5466.67 - 5700.00 & 0.11168 $\pm$ 0.00040 & 0.40 $\pm$ 0.12 & 0.29 $\pm$ 0.12 & 379 / 470 \\ 
		G750L   &                       &		 &			& \\
		5236.00 - 5673.39 & 0.11342 $\pm$ 0.00064 & 0.42 $\pm$ 0.12 & 0.28 $\pm$ 0.12 & 442 \\ 
		5673.39 - 5828.00 & 0.11178 $\pm$ 0.00068 & 0.38 $\pm$ 0.12 & 0.30 $\pm$ 0.13 & 414 \\ 
		5828.00 - 5878.00 & 0.11094 $\pm$ 0.00093 & 0.37 $\pm$ 0.12 & 0.30 $\pm$ 0.13 & 829 \\ 
		5878.00 - 5908.00 & 0.11536 $\pm$ 0.00117 & 0.38 $\pm$ 0.12 & 0.29 $\pm$ 0.13 & 723 \\
		5908.00 - 5958.00 & 0.11059 $\pm$ 0.00079 & 0.37 $\pm$ 0.12 & 0.30 $\pm$ 0.13 & 971 \\
		5958.00 - 6110.78 & 0.11034 $\pm$ 0.00060 & 0.36 $\pm$ 0.13 & 0.30 $\pm$ 0.13 & 369 \\
		6110.78 - 6329.48 & 0.11089 $\pm$ 0.00044 & 0.34 $\pm$ 0.13 & 0.30 $\pm$ 0.13 & 335 \\
		6329.48 - 6548.17 & 0.11076 $\pm$ 0.00053 & 0.32 $\pm$ 0.13 & 0.31 $\pm$ 0.13 & 353 \\
		6548.17 - 6766.87 & 0.11019 $\pm$ 0.00050 & 0.29 $\pm$ 0.13 & 0.32 $\pm$ 0.14 & 346 \\ 
		6766.87 - 6985.57 & 0.10978 $\pm$ 0.00073 & 0.30 $\pm$ 0.13 & 0.31 $\pm$ 0.14 & 490 \\ 
		6985.57 - 7204.26 & 0.11069 $\pm$ 0.00042 & 0.28 $\pm$ 0.13 & 0.30 $\pm$ 0.14 & 314 \\ 
		7204.26 - 7422.96 & 0.10919 $\pm$ 0.00072 & 0.27 $\pm$ 0.14 & 0.30 $\pm$ 0.14 & 490 \\ 
		7422.96 - 7641.65 & 0.11000 $\pm$ 0.00064 & 0.26 $\pm$ 0.14 & 0.30 $\pm$ 0.15 & 392 \\ 
		7641.65 - 7860.35 & 0.10881 $\pm$ 0.00078 & 0.24 $\pm$ 0.14 & 0.31 $\pm$ 0.15 & 519 \\ 
		7860.35 - 8079.04 & 0.11063 $\pm$ 0.00089 & 0.24 $\pm$ 0.14 & 0.30 $\pm$ 0.15 & 598 \\ 
		8079.04 - 10266.00 & 0.10990 $\pm$ 0.00079 & 0.20 $\pm$ 0.15 & 0.30 $\pm$ 0.15 & 501 \\
		\hline
	\end{tabular}
	\label{tab_bins}
\end{table*}
% Table updated after third light.

\section{Potential causes that could mimic the derived slope}
\label{sec_causes}

\subsection{Choice of limb darkening}

\cite{csizmadia2013} suggested that a difference in theoretical calculations of LDCs may lead to varying results in $\mathrm{R_p/R_s}$. To confirm that the slope in the transmission spectrum is not caused by our custom LDCs, we recomputed the transmission spectrum of WASP-76~b as previously explained, but using the LDCs for the Johnson-Cousin filters taken from \cite{claret2011} instead. As the bands in the $UBVRI$ filters are much wider than our wavelength integration bands, we used as LDCs the interpolated values corresponding to the center of each one of our wavelength bands. The remaining analysis to derive the transmission spectrum of WASP-76~b is kept as explained before. In all cases, the derived $\mathrm{R_p/R_s}$'s are consistent between each other considering their individual uncertainties at 2-$\sigma$ level.

Despite the consistency between the derived $\mathrm{R_p/R_s}$'s, to further investigate if there is any significant change in the overall slope of our transmission spectrum we compared the values derived from our LDCs (henceforth, TS1) and the ones obtained from using the linearly interpolated values of \cite{claret2011} (henceforth, TS2) by means of a Kolmogorov-Smirnov test (KST) \citep{KolmogorovS}. The null hypothesis in this test is that the $\mathrm{R_p/R_s}$'s from TS1 and TS2 are drawn from identical populations. Taking into account a value of \mbox{$\alpha = 5$\%} and a KST$_{TS1,TS2}$ = 0.38, we can not reject the null hypothesis of the two samples being drawn from the same distribution, with a 95\% confidence level. In other words, the difference in $\mathrm{R_p/R_s}$ is negligible, and the resulting transmission spectrum remains unchanged, still exhibiting the same strength in the slope, as seen in Figure~\ref{fig_transmission}, regardless the specific choice of limb darkening. 

\subsection{Stellar activity}

The effects of stellar activity over the transmission spectrum of exoplanets is one of the main limitations of the method. Both occulted and non-occulted stellar spots can have an impact on the estimates of the planet-to-star radii ratio, and in consequence on the derived results \citep[see e.g.,][]{sing11,Oshagh2014,Mallonn2018}.

Analyzing WASP's full photometric time series, \cite{west2016} did not report any variability in the system. Instead, the authors report an upper limit for the rotational modulation of WASP-76 to be as large as 1~milli-magnitude (95\% credibility interval). Using the formalism of \cite{sing11}, a stellar variability of 0.1\% would result in a modification of the transit depth by 0.1\%. In the case of WASP-76~b, the corresponding $\Delta$R$_P$/R$_S$ is below 0.0001, which is only a fraction of our typical uncertainty of the data points in the transmission spectrum and an order of magnitude smaller than the measured variation of R$_P$/R$_S$ over the optical wavelength range. In this analysis, we are interested in the chromatic difference of the modification of the transit depth, which is again about an order of magnitude smaller than the modification itself, estimated above. Therefore, we conclude that a stellar variability of amplitude 0.1\% cannot significantly affect the measured slope in the transmission spectrum.

\subsection{Impact parameter}

Following up on the discrepancies detected by \cite{Alexoudi2018} on the atmosphere of HAT-P-12b due to an inadequate choice of orbital inclination, we investigated the impact of the semi-major axis and the orbital inclination into our derived transmission spectrum. We do this by carrying out our usual MCMC runs, but in this case keeping both parameters fixed. Instead of using their literature values from \cite{seidel2019}, we used the literature values plus and minus one time their uncertainties. The parameter values and their errors can be seen in Table~\ref{tab_ait0}. After deriving the wavelength-dependent $\mathrm{R_p/R_s}$'s in the usual way, for each case we computed the Rayleigh slope as specified in Section~\ref{Rayleigh}. In all cases the derived temperatures ranged between 771 and 974 K, with corresponding $\chi^2_{red}$ values ranging between 2.17 and 3.60.

\section{Discussion}
\label{sec_discussion}

\subsection{Rayleigh scattering atmosphere}
\label{Rayleigh}

As Figure~\ref{fig_transmission} reveals, a downward slope with increasing wavelength is clearly visible. Assuming this slope to be caused by Rayleigh scattering in the atmosphere of WASP-76~b, we followed the approach proposed by \cite{lecavelier}, where the absorption depth ($AD$) of Rayleigh scattering can be expressed as follows:

\begin{equation}
    \label{eq_AD}
    AD = AD_0 \left(1 - \frac{8H}{R_p} \ln \frac{\lambda}{\lambda_0}\right),\,
\end{equation}

\noindent where $AD_0$ is the absorption depth at a reference wavelength, $\lambda_0$, and $H$ is the scale height, expressed as:

\begin{equation}
    H = kT/\mu g.
\end{equation}

\noindent In this equation, $k$ is the Boltzmann constant, $\mu$ is the mean mass of atmospheric particles, considered to be 2 times the mass of the proton, and $g$ is the surface gravity. As pointed out by \cite{lecavelier}, Equation \ref{eq_AD} is only true for $R_p/H$ between 30 and 300. For WASP-76~b $R_P/H$ was calculated to be around 90 for all wavelength bins, so we can therefore safely use Equation \ref{eq_AD}.

The fit to Equation \ref{eq_AD} was done with MCMC, considering $T$ and $AD_0$ as fitting parameters. In addition, we considered uniform priors with starting values of 2160 K \citep{west2016} for T, and 0.11034 $\mathrm{R_p/R_s}$ for $AD_0$, which is the 14$^{th}$ data point in the transmission spectrum. The result can be seen in Figure~\ref{fig_transmission} as a black line with the derived temperature of \mbox{$T$ = 866 $\pm$ 114 K} and \mbox{$AD_0$ = 0.11112 $\pm$ 0.00011}, for a reduced $\chi^2$ of \mbox{$\chi^2_{\mathrm{red}}$ = 2.17}. 

The rather large $\chi^2_{\mathrm{red}}$ value between the Rayleigh
model and the observed transmission spectrum seems also to indicate
that Rayleigh scattering alone is not responsible for the observed
spectrum. Another potential description of the measured slope might be
general Mie scattering theory instead of Rayleigh scattering,
e.g. scattering by particles larger than sub-micron size
\citep{sing13,Mallonn2016,Wakeford2015}. However, the atmospheric
retrieval, detailed in Section~\ref{sec_retrieval}, favors a clear
atmosphere and disfavors scattering haze particles.

\subsection{Atmospheric retrieval}
\label{sec_retrieval}

We performed a retrieval of the atmospheric properties of WASP-76~b from the optical transmission spectrum derived in this work, using an adaptation of the AURA retrieval code \citep{Pinhas2018,Welbanks2019}. The code consists of two main components: a forward model, which computes theoretical transmission spectra assuming a plane-parallel atmosphere in hydrostatic equilibrium, and a statistical sampling algorithm, which explores the full parameter space of possible models to carry out parameter estimation and model comparison. For the statistical inference we employ a variant of the Nested Sampling algorithm called MultiNest \citep{Skilling2004,Feroz2009}. Specifically, we use PyMultiNest \citep{Buchner2014}, an implementation of this algorithm with a Python interface.

The forward model uses a paramaterized pressure-temperature profile following the prescription of \citet{Madhusudhan2009}, consisting of six free parameters. The reference pressure $P_{\rm ref}$ at the planetary white-light radius is also taken as a free parameter. A range of opacity sources are considered: extinction from chemical species, collision-induced absorption due to H$_2$-H$_2$ and H$_2$-He interactions, and cloud/haze opacity. The mixing ratios of H$_2$ and He are determined by assuming a solar composition of $X_{\textnormal{He}} / X_{\textnormal{H}_2} = 0.18$ \citep{Asplund2009} and using the fact that the sum of all mixing ratios must equal unity. The atmospheric model is capable of incorporating a wide range of chemical species, with opacities calculated based on \citet{Gandhi2018}. Species with prominent optical cross-sections considered in the model are Na, K, Li, TiO, VO, AlO, CaO, TiH, CrH, FeH, and ScH. Line lists for these species are taken from the ExoMol database \citep{Tennyson2016}, including \citet{Burrows2005} for TiH, \citet{Burrows2002} for CrH and \citet{Lodi2015} for ScH. The mixing ratios of these atomic and molecular species are considered free parameters. We use the cloud parameterization of \citet{MacDonald2017} which introduces an additional source of opacity into the model due to clouds/hazes:

\begin{equation}
    \kappa_{\rm cloud} =
    \begin{cases}
      a \, \sigma_0 (\lambda/\lambda_0)^{\gamma}, & P < P_{\rm cloud} \\
      \infty, & P \geq P_{\rm cloud}.
    \end{cases}
\end{equation}

This prescription incorporates four additional free parameters into the model: the Rayleigh-enhancement factor $a$, the haze slope $\gamma$, the cloud top pressure $P_{\rm cloud}$ and the fractional cloud coverage $\bar{\phi}$. 

\begin{figure}
    \centering
    \includegraphics[width=\columnwidth]{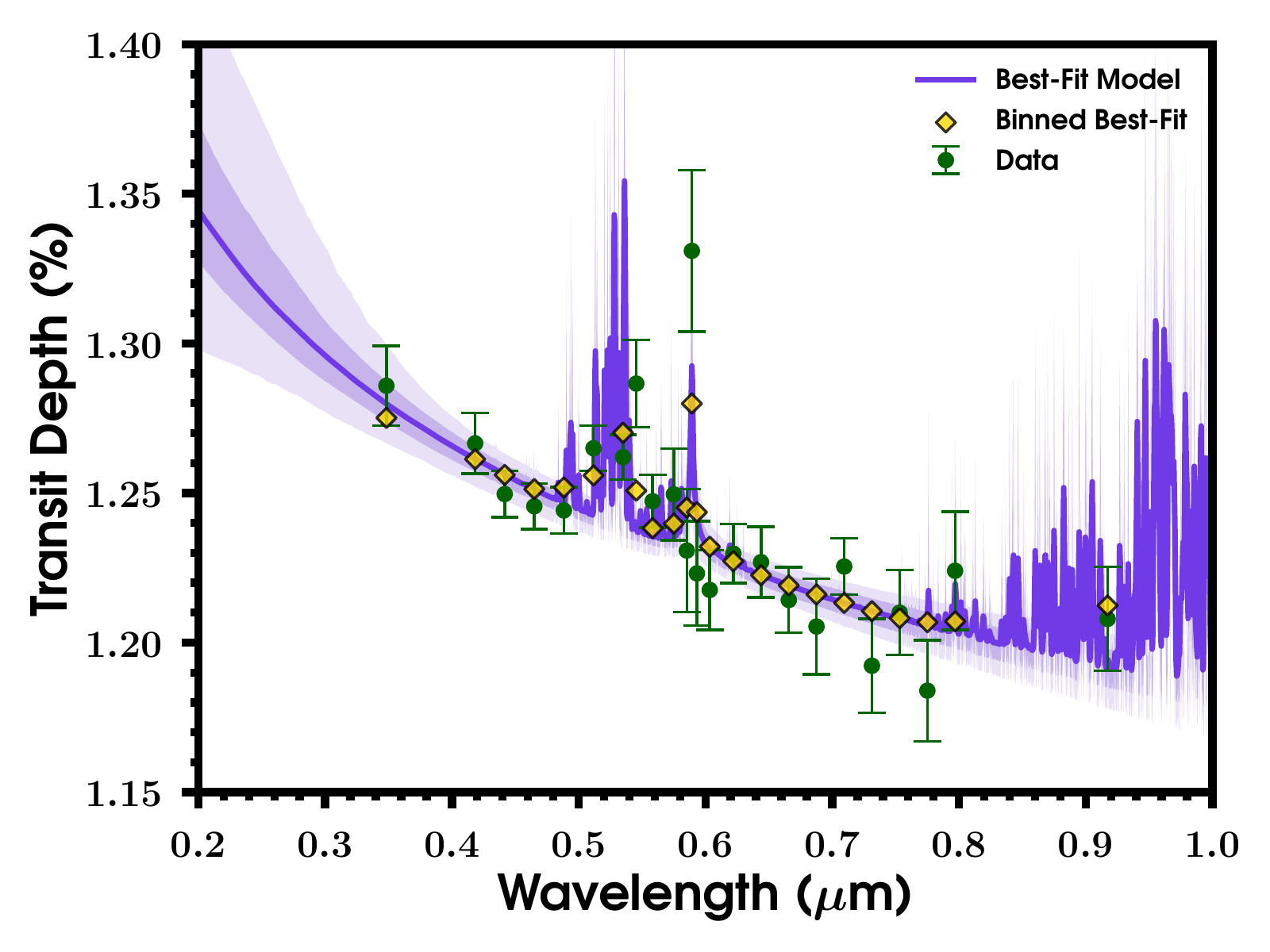}
    \caption{Retrieved best-fit transmission spectrum of WASP-76~b. The dark- and light-shaded areas represent 1- and 2-$\sigma$ contours respectively, produced by drawing a sample of 1000 spectra from the retrieved posterior probability distributions. The median best-fit spectrum has been smoothed with a Gaussian filter.}
    \label{fig:retrieval}
\end{figure}

The best-fitting retrieved spectrum is shown in Figure
\ref{fig:retrieval}. The reduced $\chi^2$ for the best-fit model is
$\chi^2_{\rm red}=1.51$. Comparing the Bayesian evidence for the
best-fitting model against a model with no Na absorption, but that is
otherwise identical, indicates a detection of Na at $2.9\sigma$
confidence. We retrieve an abundance of $\log X_{\rm
  Na}=-8.3^{+0.7}_{-0.8}$. The retrieved temperature at the top of our
atmosphere ($10^{-5}$ bar) is $2300^{+412}_{-392}$ K, consistent with
the equilibrium temperature of the planet, T$_\mathrm{eq}$ = 2160
$\pm$ 40 K for an albedo equal to zero \citep{west2016}. We also find
marginal evidence for titanium hydride, which may explain the feature
around 0.5-0.55$\mu$m. We obtain an upper limit for TiH abundance;
$\log X_{\rm TiH} \leq -8.6$. The evidence ratio of the model
including only Na and TiH versus the full set of opacities discussed
above is 5.3 in favour of the model with fewer chemical species,
indicating that other species are not detectable with the present data
\citep{Trotta2008}.

Models incorporating the cloud/haze prescription described above are only slightly preferred to cloud-free models at $1.4\sigma$ confidence. The retrieved value of $\bar{\phi} = 0.18^{+0.13}_{-0.11}$ suggests a mostly clear atmosphere. This result indicates that a Rayleigh slope in combination with several atomic or molecular species is the best explanation for the observed transmission spectrum. 

\subsection{WASP-76~b in context}

A spectral slope caused by scattering from either gas molecules or condensates has been measured in the transmission spectra of many Hot Jupiter exoplanets \citep[e.g.,][]{sing11,sing15,nikolov15,Mallonn2016,Chen2017}. Next to absorption features of alkali metals, such a slope is one of very few features in optical transmission spectra that can be used for atmospheric characterization at the current level of commonly achieved measurement precision. 

The amplitude of the optical spectral slope of Hot Jupiters has been measured to be typically of order of one to two atmospheric pressure scale heights from the near-UV to the near IR \citep{sing16}. No correlation of slope amplitude with planetary temperature and gravity has been found yet \citep{Mallonn2017}. The amplitude of the slope measured for WASP-76~b turns out to be relatively strong compared to other Hot Jupiters. It is similar in units of pressure scale height to, e.g., HAT-P-32b \citep{Mallonn2017}, and WASP-6b \citep{nikolov15}, but less steep than, e.g., HD~189733b \citep{sing11} and GJ~3470b \citep{Chen2017,Nascimbeni2013}. Among the ultra-hot planets, a variety of slopes have been found from nearly zero for WASP-19b \citep{Espinoza2019} and WASP-103b \citep{Lendl2017} to the sloped spectra of WASP-12b \citep{sing13} and WASP-76~b detected in this work. Thus, ultra-hot Jupiters seem to show a similar diversity in the amplitude of their optical spectral slope as Hot Jupiters of more moderate temperature. We want to note the result of recent studies which showed the slope to be very vulnerable to the effect of systematics, either of instrumental or astrophysical origin \citep{Mackebrandt2017,Mallonn2018,Alexoudi2018}. This observational challenge also becomes evident in numerous conflicting published optical slopes of individual targets, e.g. for WASP-19b, WASP-80b, TrES-3b, or GJ~3470b. 

The derived transmission spectrum of WASP-76~b shares similarities to
the spectrum of WASP-17b \citep{sing16}; both show a significant slope
and absorption by sodium in low spectral resolution. An atmospheric
retrieval of \cite{Barstow2017} on the WASP-17b data of \cite{sing16}
yielded a Rayleigh-scattering haze layer with a condensate top layer
slightly deeper in the atmosphere than for other planets with
scattering signature. If similar for WASP-76~b, we would expect to
find pronounced water absorption by several scale heights in the near
IR for solar composition, only weakly obscured by the relatively deep
haze layer. Indeed, the WFC3 transmission spectrum of WASP-76~b,
covering wavelengths between 1.1 and 1.6 microns, was analyzed by
\cite{tsiaras}, who found a significant water feature plus signs of
TiO/VO. Contrary to this, \cite{Fisher2018} used a non-grey cloud
model to fit the observed slope, finding a lower water abundance and
no detection of TiO/VO. While carrying out our first determination of
the wavelength-dependent transit depths, we were not aware of
WASP-76's very bright companion. After comparing the transmission
spectrum computed with and without considering the companion, we
realized the derived transmission spectrum were offset between each
other and had a different slope, making the two set of values
inconsistent even at a 3-$\sigma$ level. Both \cite{tsiaras} and
\cite{Fisher2018} do not take into account the presence of the
companion and thus neglect the effect that this can have on their
derived transmission spectrum. In consequence, we caution the reader
against making any conclusion regarding the detection inconsistency of
TiO/VO in the atmosphere of WASP-76~b, as this might be caused by the
lack of treatment of the stellar companion.

Sodium absorption at the terminator of WASP-76~b was reported by \cite{seidel2019} using high-resolution transmission spectroscopy. Qualitatively, we confirm the sodium absorption with our approach of low-resolution transmission spectroscopy. However, our value of excess absorption of about 0.15\% in a bandwidth of 20~\AA\ compared to the average of the two adjacent bands seems much stronger than the value of \cite{seidel2019} of 0.37\% in the very narrow band of 0.75~\AA. To illustrate this difference, we compare the WASP-76~b measurements to HD~189733b. The high-resolution sodium absorption strength of WASP-76~b and HD~189733b is very similar \citep{seidel2019,Wyttenbach2015}. However, our low-resolution sodium absorption for WASP-76~b in this work is one order of magnitude stronger than the low-resolution sodium absorption of HD~189733b \citep{sing16}. Thus, future follow-up measurements need to verify if the strength of sodium detected in this work is overestimated.

\section{Conclusion}
\label{sec_conclusion}

In this work, we present the derived primordial composition of the atmosphere of WASP-76~b, covering the near UV to near IR wavelengths. To do so, we analyzed in simultaneous three primary transits obtained with HST/STIS. At first, we observe a large variability in the transmission spectrum, which corresponds to approximately an extension of 3 scale heights, assuming equilibrium temperature. We analyzed some potential causes that could mimic it, including the choice of limb darkening, the impact of stellar activity and the choice of orbital parameters, with special emphasize on the inclination and the semi-major axis. Assuming that the slope is caused by Rayleigh scattering, we derived an atmospheric temperature of \mbox{866 $\pm$ 114 K}. A \mbox{$\chi^2_{red}$ = 2.17} between data and model, and a clear outlying point around sodium observed in the transmission spectrum, motivated us to carry out an atmospheric retrieval. Through it, we detect sodium at a 2.9$\sigma$ confidence, and retrieve an abundance of $\log X_{\rm Na}=-8.3^{+0.7}_{-0.8}$. The retrieved temperature of \mbox{2300$^{+412}_{-392}$ K} is consistent with the equilibrium temperature of the planet. Besides sodium, we obtain some evidence for titanium hydride, however this is still a tentative detection.

\begin{acknowledgements}

CvE and HK acknowledge funding for the Stellar Astrophysics Centre, provided by The Danish National Research Foundation (Grant DNRF106). CvE, HK and GT acknowledge support from the European Social Fund via the Lithuanian Science Council (LMTLT) grant No. 09.3.3-LMT-K-712-01-0103. This work made use of PyAstronomy\footnote{https://github.com/sczesla/PyAstronomy}. 

\end{acknowledgements}

\bibliographystyle{aa}
\bibliography{main}

\begin{thebibliography}{77}
\expandafter\ifx\csname natexlab\endcsname\relax\def\natexlab#1{#1}\fi

\bibitem[{{Alexoudi} {et~al.}(2018){Alexoudi}, {Mallonn}, {von Essen},
  {Turner}, {Keles}, {Southworth}, {Mancini}, {Ciceri}, {Granzer}, \&
  {Denker}}]{Alexoudi2018}
{Alexoudi}, X., {Mallonn}, M., {von Essen}, C., {et~al.} 2018, \aap, 620, A142

\bibitem[{{Asplund} {et~al.}(2009){Asplund}, {Grevesse}, {Sauval}, \&
  {Scott}}]{Asplund2009}
{Asplund}, M., {Grevesse}, N., {Sauval}, A.~J., \& {Scott}, P. 2009, \araa, 47,
  481

\bibitem[{{Barstow} {et~al.}(2017){Barstow}, {Aigrain}, {Irwin}, \&
  {Sing}}]{Barstow2017}
{Barstow}, J.~K., {Aigrain}, S., {Irwin}, P.~G.~J., \& {Sing}, D.~K. 2017,
  \apj, 834, 50

\bibitem[{{Boyajian} {et~al.}(2015){Boyajian}, {von Braun}, {Feiden}, {Huber},
  {Basu}, {Demarque}, {Fischer}, {Schaefer}, {Mann}, {White}, {Maestro},
  {Brewer}, {Lamell}, {Spada}, {L{\'o}pez-Morales}, {Ireland}, {Farrington},
  {van Belle}, {Kane}, {Jones}, {ten Brummelaar}, {Ciardi}, {McAlister},
  {Ridgway}, {Goldfinger}, {Turner}, \& {Sturmann}}]{Boyajian2015}
{Boyajian}, T., {von Braun}, K., {Feiden}, G.~A., {et~al.} 2015, \mnras, 447,
  846

\bibitem[{{Brown} {et~al.}(2017){Brown}, {Triaud}, {Doyle}, {Gillon}, {Lendl},
  {Anderson}, {Collier Cameron}, {H{\'e}brard}, {Hellier}, {Lovis}, {Maxted},
  {Pepe}, {Pollacco}, {Queloz}, \& {Smalley}}]{star}
{Brown}, D.~J.~A., {Triaud}, A.~H.~M.~J., {Doyle}, A.~P., {et~al.} 2017,
  \mnras, 464, 810

\bibitem[{{Brown} {et~al.}(2001){Brown}, {Charbonneau}, {Gilliland}, {Noyes},
  \& {Burrows}}]{brown2001}
{Brown}, T.~M., {Charbonneau}, D., {Gilliland}, R.~L., {Noyes}, R.~W., \&
  {Burrows}, A. 2001, \apj, 552, 699

\bibitem[{{Buchner} {et~al.}(2014){Buchner}, {Georgakakis}, {Nandra}, {Hsu},
  {Rangel}, {Brightman}, {Merloni}, {Salvato}, {Donley}, \&
  {Kocevski}}]{Buchner2014}
{Buchner}, J., {Georgakakis}, A., {Nandra}, K., {et~al.} 2014, \aap, 564, A125

\bibitem[{{Burrows} {et~al.}(2005){Burrows}, {Dulick}, {Bauschlicher},
  {Bernath}, {Ram}, {Sharp}, \& {Milsom}}]{Burrows2005}
{Burrows}, A., {Dulick}, M., {Bauschlicher}, C.~W., J., {et~al.} 2005, \apj,
  624, 988

\bibitem[{{Burrows} {et~al.}(2002){Burrows}, {Ram}, {Bernath}, {Sharp}, \&
  {Milsom}}]{Burrows2002}
{Burrows}, A., {Ram}, R.~S., {Bernath}, P., {Sharp}, C.~M., \& {Milsom}, J.~A.
  2002, \apj, 577, 986

\bibitem[{{Charbonneau} {et~al.}(2002){Charbonneau}, {Brown}, {Noyes}, \&
  {Gilliland}}]{charbonneau}
{Charbonneau}, D., {Brown}, T.~M., {Noyes}, R.~W., \& {Gilliland}, R.~L. 2002,
  \apj, 568, 377

\bibitem[{{Chen} {et~al.}(2017){Chen}, {Guenther}, {Pall{\'e}}, {Nortmann},
  {Nowak}, {Kunz}, {Parviainen}, \& {Murgas}}]{Chen2017}
{Chen}, G., {Guenther}, E.~W., {Pall{\'e}}, E., {et~al.} 2017, \aap, 600, A138

\bibitem[{{Chen} {et~al.}(2018){Chen}, {Pall{\'e}}, {Welbanks},
  {Prieto-Arranz}, {Madhusudhan}, {Gandhi}, {Casasayas-Barris}, {Murgas},
  {Nortmann}, {Crouzet}, {Parviainen}, \& {Gandolfi}}]{Chen2018}
{Chen}, G., {Pall{\'e}}, E., {Welbanks}, L., {et~al.} 2018, \aap, 616, A145

\bibitem[{{Claret}(2000)}]{claret2000}
{Claret}, A. 2000, \aap, 363, 1081

\bibitem[{{Claret} \& {Bloemen}(2011)}]{claret2011}
{Claret}, A. \& {Bloemen}, S. 2011, \aap, 529, A75

\bibitem[{{Csizmadia} {et~al.}(2013){Csizmadia}, {Pasternacki}, {Dreyer},
  {Cabrera}, {Erikson}, \& {Rauer}}]{csizmadia2013}
{Csizmadia}, S., {Pasternacki}, T., {Dreyer}, C., {et~al.} 2013, \aap, 549, A9

\bibitem[{{Damiano} {et~al.}(2017){Damiano}, {Morello}, {Tsiaras}, {Zingales},
  \& {Tinetti}}]{damiano}
{Damiano}, M., {Morello}, G., {Tsiaras}, A., {Zingales}, T., \& {Tinetti}, G.
  2017, \aj, 154, 39

\bibitem[{{Deming} {et~al.}(2013){Deming}, {Wilkins}, {McCullough}, {Burrows},
  {Fortney}, {Agol}, {Dobbs-Dixon}, {Madhusudhan}, {Crouzet}, {Desert},
  {Gilliland}, {Haynes}, {Knutson}, {Line}, {Magic}, {Mandell}, {Ranjan},
  {Charbonneau}, {Clampin}, {Seager}, \& {Showman}}]{deming}
{Deming}, D., {Wilkins}, A., {McCullough}, P., {et~al.} 2013, \apj, 774, 95

\bibitem[{{Eastman} {et~al.}(2010){Eastman}, {Siverd}, \&
  {Gaudi}}]{Eastman2010}
{Eastman}, J., {Siverd}, R., \& {Gaudi}, B.~S. 2010, \pasp, 122, 935

\bibitem[{{Espinoza} {et~al.}(2019){Espinoza}, {Rackham}, {Jord{\'a}n}, {Apai},
  {L{\'o}pez-Morales}, {Osip}, {Grimm}, {Hoeijmakers}, {Wilson}, {Bixel},
  {McGruder}, {Rodler}, {Weaver}, {Lewis}, {Fortney}, \&
  {Fraine}}]{Espinoza2019}
{Espinoza}, N., {Rackham}, B.~V., {Jord{\'a}n}, A., {et~al.} 2019, \mnras, 482,
  2065

\bibitem[{{Evans} {et~al.}(2018){Evans}, {Sing}, {Goyal}, {Nikolov}, {Marley},
  {Zahnle}, {Henry}, {Barstow}, {Alam}, {Sanz-Forcada}, {Kataria}, {Lewis},
  {Lavvas}, {Ballester}, {Ben-Jaffel}, {Blumenthal}, {Bourrier}, {Drummond},
  {Garc{\'{\i}}a Mu{\~n}oz}, {L{\'o}pez-Morales}, {Tremblin}, {Ehrenreich},
  {Wakeford}, {Buchhave}, {Lecavelier des Etangs}, {H{\'e}brard}, \&
  {Williamson}}]{evans2018}
{Evans}, T.~M., {Sing}, D.~K., {Goyal}, J.~M., {et~al.} 2018, \aj, 156, 283

\bibitem[{{Feroz} {et~al.}(2009){Feroz}, {Hobson}, \& {Bridges}}]{Feroz2009}
{Feroz}, F., {Hobson}, M.~P., \& {Bridges}, M. 2009, \mnras, 398, 1601

\bibitem[{{Fischer} {et~al.}(2016){Fischer}, {Knutson}, {Sing}, {Henry},
  {Williamson}, {Fortney}, {Burrows}, {Kataria}, {Nikolov}, {Showman},
  {Ballester}, {D{\'e}sert}, {Aigrain}, {Deming}, {Lecavelier des Etangs}, \&
  {Vidal-Madjar}}]{fischer}
{Fischer}, P.~D., {Knutson}, H.~A., {Sing}, D.~K., {et~al.} 2016, The
  Astrophysical Journal, 827, 19

\bibitem[{{Fisher} \& {Heng}(2018)}]{Fisher2018}
{Fisher}, C. \& {Heng}, K. 2018, \mnras, 481, 4698

\bibitem[{{Gandhi} \& {Madhusudhan}(2018)}]{Gandhi2018}
{Gandhi}, S. \& {Madhusudhan}, N. 2018, \mnras, 474, 271

\bibitem[{{Ginski} {et~al.}(2016){Ginski}, {Mugrauer}, {Seeliger}, {Buder},
  {Errmann}, {Avenhaus}, {Mouillet}, {Maire}, \& {Raetz}}]{Ginski2016}
{Ginski}, C., {Mugrauer}, M., {Seeliger}, M., {et~al.} 2016, \mnras, 457, 2173

\bibitem[{Göttingen(2018)}]{phoenix}
Göttingen, G.-A.-U. 2018, Göttingen Spectral Library by PHOENIX

\bibitem[{{Hartman} {et~al.}(2016){Hartman}, {Bakos}, {Bhatti}, {Penev},
  {Bieryla}, {Latham}, {Kov{\'a}cs}, {Torres}, {Csubry}, {de Val-Borro},
  {Buchhave}, {Kov{\'a}cs}, {Quinn}, {Howard}, {Isaacson}, {Fulton}, {Everett},
  {Esquerdo}, {B{\'e}ky}, {Szklenar}, {Falco}, {Santerne}, {Boisse},
  {H{\'e}brard}, {Burrows}, {L{\'a}z{\'a}r}, {Papp}, \&
  {S{\'a}ri}}]{Hartman2016}
{Hartman}, J.~D., {Bakos}, G.~{\'A}., {Bhatti}, W., {et~al.} 2016, \aj, 152,
  182

\bibitem[{{Hoeijmakers} {et~al.}(2019){Hoeijmakers}, {Ehrenreich}, {Kitzmann},
  {Allart}, {Grimm}, {Seidel}, {Wyttenbach}, {Pino}, {Nielsen}, {Fisher},
  {Rimmer}, {Bourrier}, {Cegla}, {Lavie}, {Lovis}, {Patzer}, {Stock}, {Pepe},
  \& {Heng}}]{Hoeijmakers2019}
{Hoeijmakers}, H.~J., {Ehrenreich}, D., {Kitzmann}, D., {et~al.} 2019, \aap,
  627, A165

\bibitem[{{Huitson} {et~al.}(2013){Huitson}, {Sing}, {Pont}, {Fortney},
  {Burrows}, {Wilson}, {Ballester}, {Nikolov}, {Gibson}, {Deming}, {Aigrain},
  {Evans}, {Henry}, {Lecavelier des Etangs}, {Showman}, {Vidal-Madjar}, \&
  {Zahnle}}]{huitson}
{Huitson}, C.~M., {Sing}, D.~K., {Pont}, F., {et~al.} 2013, \mnras, 434, 3252

\bibitem[{{Jones} {et~al.}(2001){Jones}, {Oliphant}, {Peterson},
  {et~al.}}]{Jones2001}
{Jones}, E., {Oliphant}, T., {Peterson}, P., {et~al.} 2001, {SciPy}: Open
  source scientific tools for {Python}, \url{http://www.scipy.org}

\bibitem[{Karson(1968)}]{KolmogorovS}
Karson, M. 1968, Journal of the American Statistical Association, 63, 1047

\bibitem[{Kass \& Raftery(1995)}]{Kass1995}
Kass, R.~E. \& Raftery, A.~E. 1995, Journal of the American Statistical
  Association, 90, 773

\bibitem[{{Kervella} {et~al.}(2017){Kervella}, {Bigot}, {Gallenne}, \&
  {Th{\'e}venin}}]{Kervella2017}
{Kervella}, P., {Bigot}, L., {Gallenne}, A., \& {Th{\'e}venin}, F. 2017, \aap,
  597, A137

\bibitem[{{Kirk} {et~al.}(2017){Kirk}, {Wheatley}, {Louden}, {Doyle},
  {Skillen}, {McCormac}, {Irwin}, \& {Karjalainen}}]{Kirk2017}
{Kirk}, J., {Wheatley}, P.~J., {Louden}, T., {et~al.} 2017, \mnras, 468, 3907

\bibitem[{{Lecavelier Des Etangs} {et~al.}(2008){Lecavelier Des Etangs},
  {Vidal-Madjar}, {D{\'e}sert}, \& {Sing}}]{lecavelier}
{Lecavelier Des Etangs}, A., {Vidal-Madjar}, A., {D{\'e}sert}, J.-M., \&
  {Sing}, D. 2008, \aap, 485, 865

\bibitem[{{Lendl} {et~al.}(2017){Lendl}, {Cubillos}, {Hagelberg}, {M{\"u}ller},
  {Juvan}, \& {Fossati}}]{Lendl2017}
{Lendl}, M., {Cubillos}, P.~E., {Hagelberg}, J., {et~al.} 2017, \aap, 606, A18

\bibitem[{{Lendl} {et~al.}(2016){Lendl}, {Delrez}, {Gillon}, {Madhusudhan},
  {Jehin}, {Queloz}, {Anderson}, {Demory}, \& {Hellier}}]{Lendl2016}
{Lendl}, M., {Delrez}, L., {Gillon}, M., {et~al.} 2016, \aap, 587, A67

\bibitem[{{Lodi} {et~al.}(2015){Lodi}, {Yurchenko}, \& {Tennyson}}]{Lodi2015}
{Lodi}, L., {Yurchenko}, S.~N., \& {Tennyson}, J. 2015, Molecular Physics, 113,
  1998

\bibitem[{{MacDonald} \& {Madhusudhan}(2017)}]{MacDonald2017}
{MacDonald}, R.~J. \& {Madhusudhan}, N. 2017, \mnras, 469, 1979

\bibitem[{{Mackebrandt} {et~al.}(2017){Mackebrandt}, {Mallonn}, {Ohlert},
  {Granzer}, {Lalitha}, {Garc{\'\i}a Mu{\~n}oz}, {Gibson}, {Lee}, {Sozzetti},
  {Turner}, {Va{\v{n}}ko}, \& {Strassmeier}}]{Mackebrandt2017}
{Mackebrandt}, F., {Mallonn}, M., {Ohlert}, J.~M., {et~al.} 2017, \aap, 608,
  A26

\bibitem[{{Madhusudhan} \& {Seager}(2009)}]{Madhusudhan2009}
{Madhusudhan}, N. \& {Seager}, S. 2009, \apj, 707, 24

\bibitem[{{Mallonn} {et~al.}(2018){Mallonn}, {Herrero}, {Juvan}, {von Essen},
  {Rosich}, {Ribas}, {Granzer}, {Alexoudi}, \& {Strassmeier}}]{Mallonn2018}
{Mallonn}, M., {Herrero}, E., {Juvan}, I.~G., {et~al.} 2018, \aap, 614, A35

\bibitem[{{Mallonn} \& {Strassmeier}(2016)}]{Mallonn2016}
{Mallonn}, M. \& {Strassmeier}, K.~G. 2016, \aap, 590, A100

\bibitem[{{Mallonn} \& {Wakeford}(2017)}]{Mallonn2017}
{Mallonn}, M. \& {Wakeford}, H.~R. 2017, Astronomische Nachrichten, 338, 773

\bibitem[{{Mandel} \& {Agol}(2002)}]{mandel}
{Mandel}, K. \& {Agol}, E. 2002, \apjl, 580, L171

\bibitem[{{Nascimbeni} {et~al.}(2013){Nascimbeni}, {Piotto}, {Pagano}, {Scand
  ariato}, {Sani}, \& {Fumana}}]{Nascimbeni2013}
{Nascimbeni}, V., {Piotto}, G., {Pagano}, I., {et~al.} 2013, \aap, 559, A32

\bibitem[{{Ngo} {et~al.}(2016){Ngo}, {Knutson}, {Hinkley}, {Bryan}, {Crepp},
  {Batygin}, {Crossfield}, {Hansen}, {Howard}, {Johnson}, {Mawet}, {Morton},
  {Muirhead}, \& {Wang}}]{Ngo2016}
{Ngo}, H., {Knutson}, H.~A., {Hinkley}, S., {et~al.} 2016, \apj, 827, 8

\bibitem[{{Nikolov} {et~al.}(2015){Nikolov}, {Sing}, {Burrows}, {Fortney},
  {Henry}, {Pont}, {Ballester}, {Aigrain}, {Wilson}, {Huitson}, {Gibson},
  {D{\'e}sert}, {Lecavelier Des Etangs}, {Showman}, {Vidal-Madjar}, {Wakeford},
  \& {Zahnle}}]{nikolov15}
{Nikolov}, N., {Sing}, D.~K., {Burrows}, A.~S., {et~al.} 2015, \mnras, 447, 463

\bibitem[{{Nikolov} {et~al.}(2014){Nikolov}, {Sing}, {Pont}, {Burrows},
  {Fortney}, {Ballester}, {Evans}, {Huitson}, {Wakeford}, {Wilson}, {Aigrain},
  {Deming}, {Gibson}, {Henry}, {Knutson}, {Lecavelier des Etangs}, {Showman},
  {Vidal-Madjar}, \& {Zahnle}}]{nikolov14}
{Nikolov}, N., {Sing}, D.~K., {Pont}, F., {et~al.} 2014, \mnras, 437, 46

\bibitem[{{Oshagh} {et~al.}(2014){Oshagh}, {Santos}, {Ehrenreich},
  {Haghighipour}, {Figueira}, {Santerne}, \& {Montalto}}]{Oshagh2014}
{Oshagh}, M., {Santos}, N.~C., {Ehrenreich}, D., {et~al.} 2014, \aap, 568, A99

\bibitem[{{Parmentier} {et~al.}(2018){Parmentier}, {Line}, {Bean}, {Mansfield},
  {Kreidberg}, {Lupu}, {Visscher}, {D{\'e}sert}, {Fortney}, {Deleuil},
  {Arcangeli}, {Showman}, \& {Marley}}]{parmentier2018}
{Parmentier}, V., {Line}, M.~R., {Bean}, J.~L., {et~al.} 2018, \aap, 617, A110

\bibitem[{{Patil} {et~al.}(2010){Patil}, {Huard}, \& {Fonnesbeck}}]{Patil2010}
{Patil}, A., {Huard}, D., \& {Fonnesbeck}, C.~J. 2010, Journal of Statistical
  Software, 35, 1

\bibitem[{{Pinhas} {et~al.}(2018){Pinhas}, {Rackham}, {Madhusudhan}, \&
  {Apai}}]{Pinhas2018}
{Pinhas}, A., {Rackham}, B.~V., {Madhusudhan}, N., \& {Apai}, D. 2018, \mnras,
  480, 5314

\bibitem[{{Sedaghati} {et~al.}(2017){Sedaghati}, {Boffin}, {MacDonald},
  {Gandhi}, {Madhusudhan}, {Gibson}, {Oshagh}, {Claret}, \&
  {Rauer}}]{Sedaghati2017}
{Sedaghati}, E., {Boffin}, H. M.~J., {MacDonald}, R.~J., {et~al.} 2017, \nat,
  549, 238

\bibitem[{{Seidel} {et~al.}(2019){Seidel}, {Ehrenreich}, {Wyttenbach},
  {Allart}, {Lendl}, {Pino}, {Bourrier}, {Cegla}, {Lovis}, {Barrado},
  {Bayliss}, {Astudillo-Defru}, {Deline}, {Fisher}, {Heng}, {Joseph}, {Lavie},
  {Melo}, {Pepe}, {S{\'e}gransan}, \& {Udry}}]{seidel2019}
{Seidel}, J.~V., {Ehrenreich}, D., {Wyttenbach}, A., {et~al.} 2019, \aap, 623,
  A166

\bibitem[{{Sing} {et~al.}(2016){Sing}, {Fortney}, {Nikolov}, {Wakeford},
  {Kataria}, {Evans}, {Aigrain}, {Ballester}, {Burrows}, {Deming},
  {D{\'e}sert}, {Gibson}, {Henry}, {Huitson}, {Knutson}, {Lecavelier Des
  Etangs}, {Pont}, {Showman}, {Vidal-Madjar}, {Williamson}, \&
  {Wilson}}]{sing16}
{Sing}, D.~K., {Fortney}, J.~J., {Nikolov}, N., {et~al.} 2016, \nat, 529, 59

\bibitem[{{Sing} {et~al.}(2013){Sing}, {Lecavelier des Etangs}, {Fortney},
  {Burrows}, {Pont}, {Wakeford}, {Ballester}, {Nikolov}, {Henry}, {Aigrain},
  {Deming}, {Evans}, {Gibson}, {Huitson}, {Knutson}, {Showman}, {Vidal-Madjar},
  {Wilson}, {Williamson}, \& {Zahnle}}]{sing13}
{Sing}, D.~K., {Lecavelier des Etangs}, A., {Fortney}, J.~J., {et~al.} 2013,
  \mnras, 436, 2956

\bibitem[{{Sing} {et~al.}(2011){Sing}, {Pont}, {Aigrain}, {Charbonneau},
  {D{\'e}sert}, {Gibson}, {Gilliland}, {Hayek}, {Henry}, {Knutson}, {Lecavelier
  Des Etangs}, {Mazeh}, \& {Shporer}}]{sing11}
{Sing}, D.~K., {Pont}, F., {Aigrain}, S., {et~al.} 2011, \mnras, 416, 1443

\bibitem[{{Sing} {et~al.}(2015){Sing}, {Wakeford}, {Showman}, {Nikolov},
  {Fortney}, {Burrows}, {Ballester}, {Deming}, {Aigrain}, {D{\'e}sert},
  {Gibson}, {Henry}, {Knutson}, {Lecavelier des Etangs}, {Pont},
  {Vidal-Madjar}, {Williamson}, \& {Wilson}}]{sing15}
{Sing}, D.~K., {Wakeford}, H.~R., {Showman}, A.~P., {et~al.} 2015, \mnras, 446,
  2428

\bibitem[{{Skilling}(2004)}]{Skilling2004}
{Skilling}, J. 2004, in American Institute of Physics Conference Series, ed.
  R.~{Fischer}, R.~{Preuss}, \& U.~V. {Toussaint}, Vol. 735, 395--405

\bibitem[{{Sotzen} {et~al.}(2020){Sotzen}, {Stevenson}, {Sing}, {Kilpatrick},
  {Wakeford}, {Filippazzo}, {Lewis}, {H{\"o}rst}, {L{\'o}pez-Morales}, {Henry},
  {Buchhave}, {Ehrenreich}, {Fraine}, {Garc{\'\i}a Mu{\~n}oz}, {Jayaraman},
  {Lavvas}, {Lecavelier des Etangs}, {Marley}, {Nikolov}, {Rathcke}, \&
  {Sanz-Forcada}}]{Sotzen2020}
{Sotzen}, K.~S., {Stevenson}, K.~B., {Sing}, D.~K., {et~al.} 2020, \aj, 159, 5

\bibitem[{{Southworth} \& {Evans}(2016)}]{Southworth2016}
{Southworth}, J. \& {Evans}, D.~F. 2016, \mnras, 463, 37

\bibitem[{{Stevenson} {et~al.}(2014){Stevenson}, {Bean}, {Seifahrt},
  {D{\'e}sert}, {Madhusudhan}, {Bergmann}, {Kreidberg}, \&
  {Homeier}}]{Stevenson2014}
{Stevenson}, K.~B., {Bean}, J.~L., {Seifahrt}, A., {et~al.} 2014, \aj, 147, 161

\bibitem[{{Tennyson} {et~al.}(2016){Tennyson}, {Yurchenko}, {Al-Refaie},
  {Barton}, {Chubb}, {Coles}, {Diamantopoulou}, {Gorman}, {Hill}, {Lam},
  {Lodi}, {McKemmish}, {Na}, {Owens}, {Polyansky}, {Rivlin}, {Sousa-Silva},
  {Underwood}, {Yachmenev}, \& {Zak}}]{Tennyson2016}
{Tennyson}, J., {Yurchenko}, S.~N., {Al-Refaie}, A.~F., {et~al.} 2016, Journal
  of Molecular Spectroscopy, 327, 73

\bibitem[{{Trotta}(2008)}]{Trotta2008}
{Trotta}, R. 2008, Contemporary Physics, 49, 71

\bibitem[{{Tsiaras} {et~al.}(2018){Tsiaras}, {Waldmann}, {Zingales},
  {Rocchetto}, {Morello}, {Damiano}, {Karpouzas}, {Tinetti}, {McKemmish},
  {Tennyson}, \& {Yurchenko}}]{tsiaras}
{Tsiaras}, A., {Waldmann}, I.~P., {Zingales}, T., {et~al.} 2018, \aj, 155, 156

\bibitem[{{von Essen} {et~al.}(2017){von Essen}, {Cellone}, {Mallonn},
  {Albrecht}, {Micul{\'a}n}, \& {M{\"u}ller}}]{vonessen2017}
{von Essen}, C., {Cellone}, S., {Mallonn}, M., {et~al.} 2017, \aap, 603, A20

\bibitem[{{von Essen} {et~al.}(2019){von Essen}, {Mallonn}, {Welbanks},
  {Madhusudhan}, {Pinhas}, {Bouy}, \& {Weis Hansen}}]{wasp33b}
{von Essen}, C., {Mallonn}, M., {Welbanks}, L., {et~al.} 2019, \aap, 622, A71

\bibitem[{{von Essen} {et~al.}(2013){von Essen}, {Schr{\"o}ter}, {Agol}, \&
  {Schmitt}}]{corrnoise}
{von Essen}, C., {Schr{\"o}ter}, S., {Agol}, E., \& {Schmitt}, J.~H.~M.~M.
  2013, \aap, 555, A92

\bibitem[{{{\v{Z}}{\'a}k} {et~al.}(2019){{\v{Z}}{\'a}k}, {Kab{\'a}th},
  {Boffin}, {Ivanov}, \& {Skarka}}]{Zak2019}
{{\v{Z}}{\'a}k}, J., {Kab{\'a}th}, P., {Boffin}, H. M.~J., {Ivanov}, V.~D., \&
  {Skarka}, M. 2019, \aj, 158, 120

\bibitem[{{Wakeford} \& {Sing}(2015)}]{Wakeford2015}
{Wakeford}, H.~R. \& {Sing}, D.~K. 2015, \aap, 573, A122

\bibitem[{{Welbanks} \& {Madhusudhan}(2019)}]{Welbanks2019}
{Welbanks}, L. \& {Madhusudhan}, N. 2019, \aj, 157, 206

\bibitem[{{Wells} \& {Bell}(1994)}]{wells1994}
{Wells}, L.~A. \& {Bell}, D.~J. 1994, {Cleaning Images of Bad Pixels and Cosmic
  Rays Using IRAF}

\bibitem[{{West} {et~al.}(2016){West}, {Hellier}, {Almenara}, {Anderson},
  {Barros}, {Bouchy}, {Brown}, {Collier Cameron}, {Deleuil}, {Delrez}, {Doyle},
  {Faedi}, {Fumel}, {Gillon}, {G{\'o}mez Maqueo Chew}, {H{\'e}brard}, {Jehin},
  {Lendl}, {Maxted}, {Pepe}, {Pollacco}, {Queloz}, {S{\'e}gransan}, {Smalley},
  {Smith}, {Southworth}, {Triaud}, \& {Udry}}]{west2016}
{West}, R.~G., {Hellier}, C., {Almenara}, J.~M., {et~al.} 2016, \aap, 585, A126

\bibitem[{{White} {et~al.}(2013){White}, {Huber}, {Maestro}, {Bedding},
  {Ireland}, {Baron}, {Boyajian}, {Che}, {Monnier}, {Pope}, {Roettenbacher},
  {Stello}, {Tuthill}, {Farrington}, {Goldfinger}, {McAlister}, {Schaefer},
  {Sturmann}, {Sturmann}, {ten Brummelaar}, \& {Turner}}]{White2013}
{White}, T.~R., {Huber}, D., {Maestro}, V., {et~al.} 2013, \mnras, 433, 1262

\bibitem[{{W{\"o}llert} \& {Brandner}(2015)}]{Wolltert2015}
{W{\"o}llert}, M. \& {Brandner}, W. 2015, \aap, 579, A129

\bibitem[{{Wyttenbach} {et~al.}(2015){Wyttenbach}, {Ehrenreich}, {Lovis},
  {Udry}, \& {Pepe}}]{Wyttenbach2015}
{Wyttenbach}, A., {Ehrenreich}, D., {Lovis}, C., {Udry}, S., \& {Pepe}, F.
  2015, \aap, 577, A62

\end{thebibliography}

\begin{appendix}

\section{Posterior distributions for the white light curves}
\label{Appendix}

\begin{figure*}
    \centering
    \includegraphics[width=\textwidth]{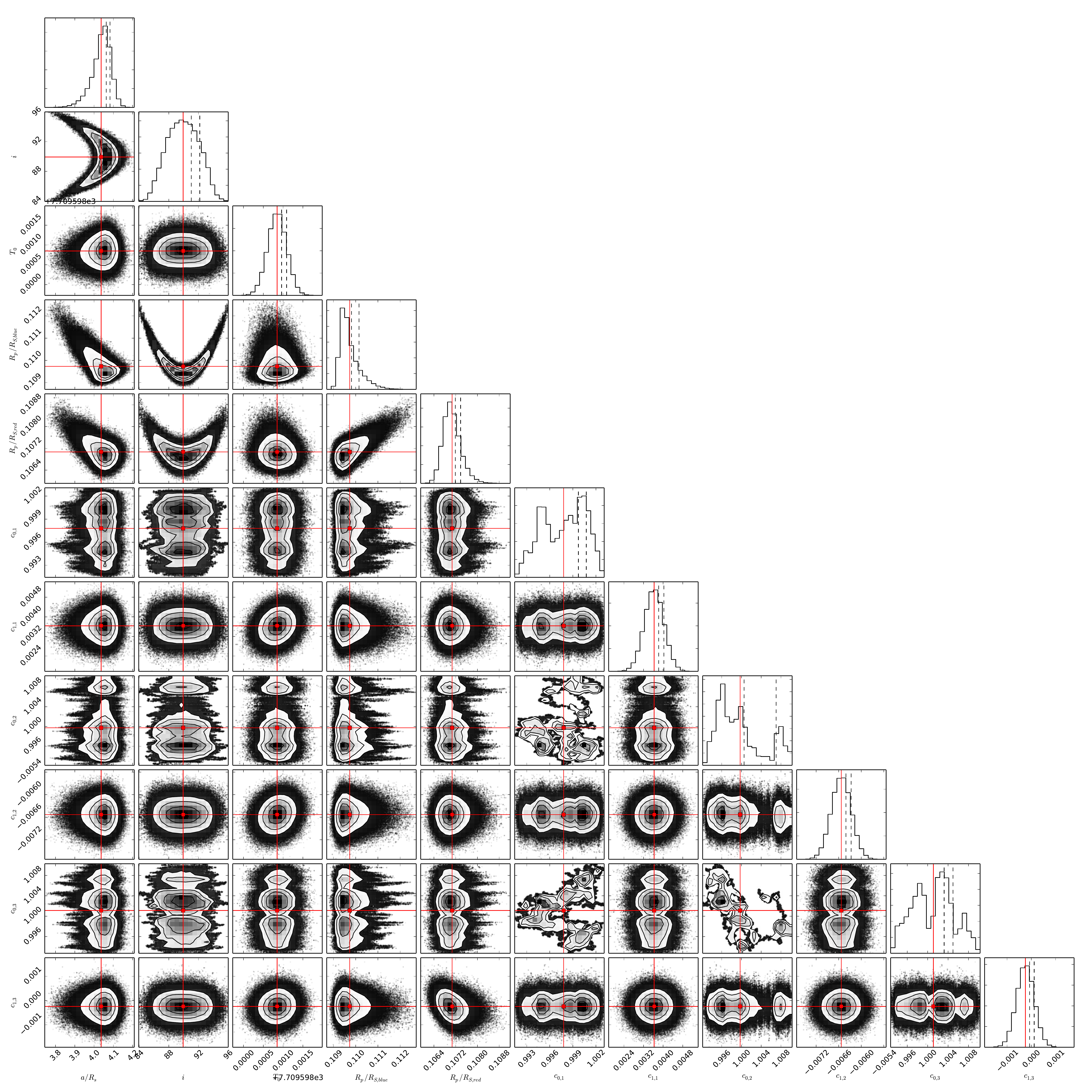}
    \caption{\label{fig:triangle1} Posterior distributions and correlations between the transit parameters, a/R$_S$, i, T$_0$ and R$_P$/R$_S$ and the coefficients for the linear slope, c$_0$ and c$_1$, for the white light curves. Subindex 1, 2 and 3 correspond to each one of the three transits, following the order of Figure~\ref{fig_white}. Red points correspond to the best-fit parameters and shaded gray to white areas correspond to 1, 2, and 3-$\sigma$ uncertainty regions.}
\end{figure*}

\begin{figure*}
    \centering
    \includegraphics[width=\textwidth]{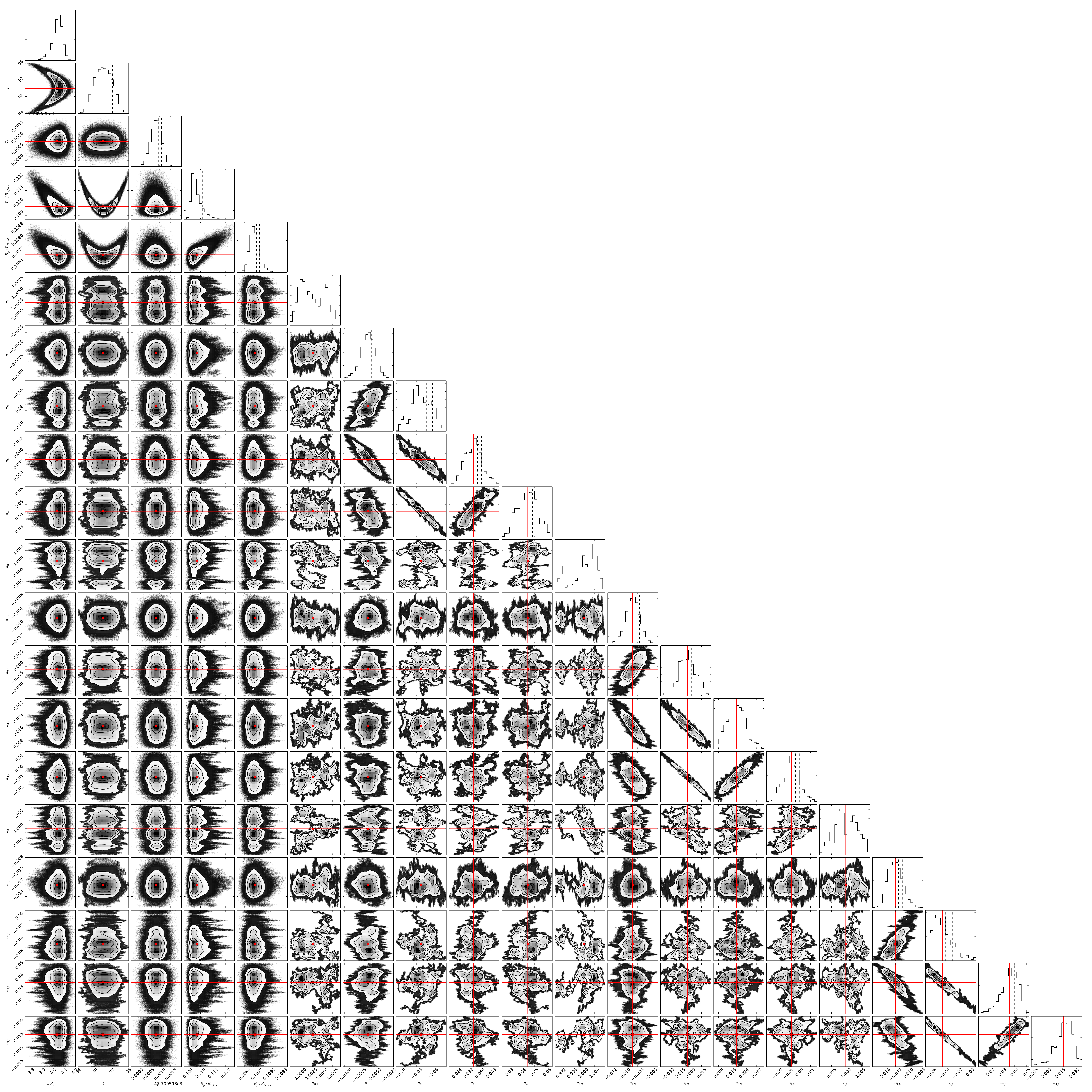}
    \caption{\label{fig:triangle2} Same as Figure~\ref{fig:triangle1}, but considering the coefficients for the fourth degree polynomial, namely a$_0$, a$_1$, a$_2$, a$_3$ and a$_4$.}
\end{figure*}

\end{appendix}

\end{document}